\PassOptionsToPackage{noend}{algpseudocode}
\documentclass[sigconf]{acmart}
\copyrightyear{2024}
\acmYear{2024}
\setcopyright{rightsretained}
\acmConference[CHI '24]{Proceedings of the CHI Conference on Human Factors
in Computing Systems}{May 11--16, 2024}{Honolulu, HI, USA}
\acmBooktitle{Proceedings of the CHI Conference on Human Factors in
Computing Systems (CHI '24), May 11--16, 2024, Honolulu, HI, USA}
\acmDOI{10.1145/3613904.3642636}
\acmISBN{979-8-4007-0330-0/24/05}

\usepackage[utf8]{inputenc}
\usepackage{enumitem}
\usepackage{subfig}
\usepackage{url}
\usepackage{multirow}
\usepackage{graphicx}
\usepackage{booktabs}
\usepackage{tabularx}
\newcolumntype{b}{X}
\newcolumntype{s}{>{\hsize=.25\hsize}X}

\usepackage{xcolor}
\colorlet{punct}{red!60!black}
\definecolor{background}{HTML}{EEEEEE}
\definecolor{delim}{RGB}{20,105,176}
\colorlet{numb}{magenta!60!black}

\usepackage{researchpack}
\algrenewcommand{\algorithmicrequire}{\textbf{Precondition:}}
\algrenewcommand{\algorithmicensure}{\textbf{Output:}}

\definecolor{BLUE}{rgb}{0,0,1}

\begin{document}

\title[Artful Path to Healing]{Artful Path to Healing: Using Machine Learning for Visual Art Recommendation to Prevent and Reduce Post-Intensive Care Syndrome (PICS)}

\author{Bereket A. Yilma}
\affiliation{%
  \institution{University of Luxembourg}
  \country{Luxembourg}
}

\author{Chan Mi Kim}
\affiliation{%
  \institution{University of Twente}
  \country{The Netherlands}
}

\author{Gerald C. Cupchik}
\affiliation{%
  \institution{University of Toronto}
  \country{Canada}
}
\author{Luis A. Leiva}
\affiliation{%
  \institution{University of Luxembourg}
  \country{Luxembourg}
}

\renewcommand{\shortauthors}{Yilma et al.}

\begin{abstract} 
Staying in the intensive care unit (ICU) is often traumatic, leading to post-intensive care syndrome (PICS), which encompasses physical, psychological, and cognitive impairments. Currently, there are limited interventions available for PICS. Studies indicate that exposure to visual art may help address the psychological aspects of PICS and be more effective if it is personalized. We develop Machine Learning-based Visual Art Recommendation Systems (VA RecSys) to enable personalized therapeutic visual art experiences for post-ICU patients. We investigate four state-of-the-art VA RecSys engines, evaluating the relevance of their recommendations for therapeutic purposes compared to expert-curated recommendations. We conduct an expert pilot test and a large-scale user study (n=150) to assess the appropriateness and effectiveness of these recommendations. Our results suggest all recommendations enhance temporal affective states. Visual and multimodal VA RecSys engines compare favourably with expert-curated recommendations, indicating their potential to  support the delivery of personalized art therapy for PICS prevention and treatment.
\end{abstract}

\begin{teaserfigure}
\includegraphics[width=\textwidth]{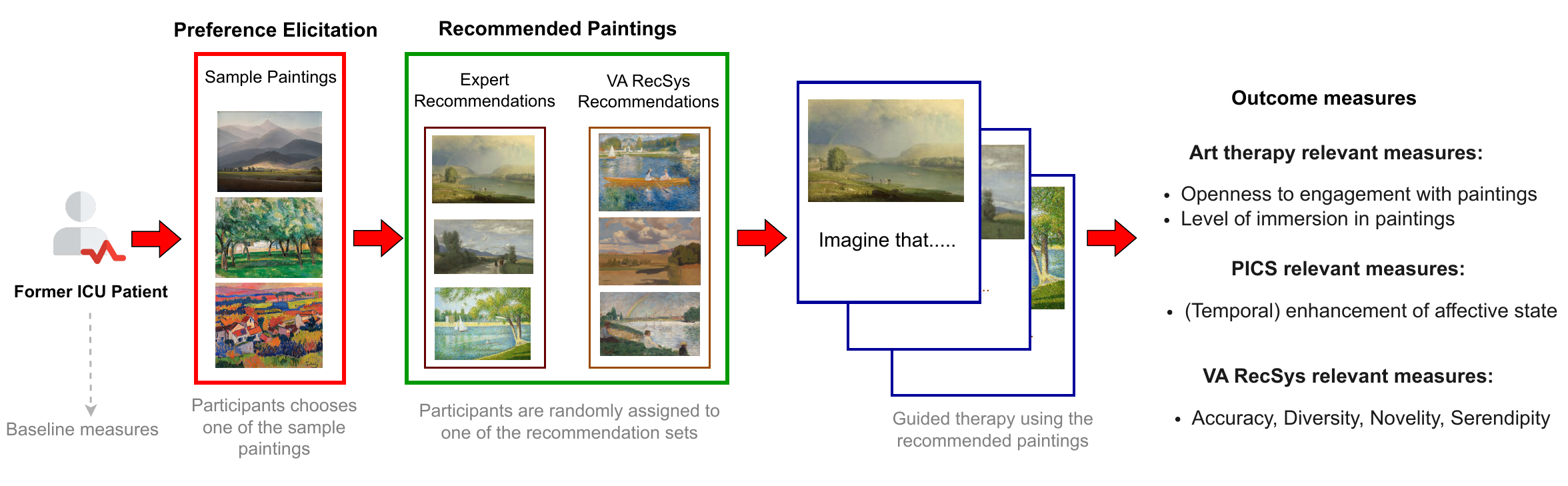}
\caption{Overview of our art therapy approach for PICS prevention and treatment.}
\label{fig:LDA_plate}
\end{teaserfigure}



\maketitle

\clearpage

\section{Introduction}

Patients in the intensive care unit (ICU)  generally undergo stressful and traumatic experiences stemming from critical illness, medical procedures, pain, and a hostile environment~\cite{egerod2015patient}. Even after ICU discharge, these patients are vulnerable and at a high risk of readmission to the hospital and the ICU~\cite{van2019dutch,shankar2020rate}. Post-ICU patients often suffer from post-intensive care syndrome (PICS) which refers to ``new or worsening impairments in physical, cognitive, or mental health arising after critical illness and persisting beyond acute care hospitalization''~\cite{needham2012improving}. PICS is quite common, affecting up to 75\% of patients discharged from the ICU~\cite{needham2013physical, rawal2017post,pandharipande2013long,davydow2013hospital} and reduced quality of life and hindrances in reintegrating into society of post-ICU patients~\cite{dowdy2005quality,griffiths2013exploration}. Psychological aspects of PICS include depression, anxiety disorders, and post-traumatic stress disorder (PTSD)~\cite{inoue2019post}. While there is growing interest in their prevention and treatment, there exist limited interventions available, such as ICU follow-up clinics~\cite{cuthbertson2009practical,inoue2019post} and ICU diaries~\cite{blair2017improving,inoue2019post}, and there is a need for more diverse and effective approaches. 

Visual art has been widely utilized to promote psychological well-being in clinical environments.  While art therapy is commonly understood as a form involving creative activities, in this paper, we use 'art therapy' as an umbrella term where art serves as a medium for therapeutic benefits  \cite{dalley2008art,geue2010overview}. This includes various forms, such as engaging with existing artwork to stimulate emotions and self-reflections. Art therapy, for example, has been employed as a method to address various forms of psychological disorders, including depression~\cite{bar2007art,geue2010overview}, anxiety~\cite{geue2010overview}, and PTSD~\cite{spiegel2006art}. This approach leverages the unique characteristics of visual art, such as diverse styles that activate interpretation and imagination, and the ability to stimulate the expression of memories and specific emotions~\cite{spiegel2006art}, which have demonstrated effectiveness in addressing these psychological disorders~\cite{bar2007art,geue2010overview,spiegel2006art}. In the context of hospitals in general, as well as critical care settings such as the ICU, the use of visual art as a positive distraction has also demonstrated its effectiveness in reducing stress, anxiety, and pain perception~\cite{John2003healing,nanda2008undertaking}. This body of evidence showcases the potential of visual art as an intervention for addressing the psychological aspects of PICS. 

Previous studies have suggested the importance of personalization in providing positive distractions to enhance effectiveness~\cite{sharda2019bach} while minimizing potential side effects~\cite{newbold2017using}. This indicates that for visual art to serve as a positive distraction, it is crucial that it resonates with the patient's emotional needs, emphasizing the significance of selecting appropriate art for each patient. Furthermore, to achieve a prolonged effect of positive distraction, a continuous supply of personalized art is necessary, which entails a large number of artworks.

In light of these challenges, it becomes evident that embracing a personalised methodology for facilitating the selection of paintings in art therapy is not merely advantageous, but of paramount importance. The intersection of personalised medicine and art therapy holds the potential to revolutionise the landscape of PICS treatment. Particularly recent advances in Machine Learning-based Visual Art Recommendation Systems (VA RecSys) hold great potential to open a novel avenue for tackling the challenges of artwork selection in a personalised and nuanced manner to be used for art therapy of post-ICU patients and beyond. By integrating these systems, we can bridge the gap between the vast universe of artworks and the unique emotional needs of each patient, thereby supporting experts in the selection process of curated artworks that resonate with the penitent's distinct cognitive and emotional requirements.

In this paper, we set out to explore the potential benefits of integrating Machine Learning (ML) based VA RecSys within the framework of PICS treatment using art therapy. To the best of our knowledge, there are no prior works leveraging VA RecSys in a therapeutic context. Therefore, we formulate the following research question: \textbf{Can VA RecSys algorithms support the  psychological well-being}  of post-ICU patients  through personalized art therapy?

In pursuit of this investigation, we propose approaches to integrate state-of-the-art VA RecSys engines within the context of PICS prevention and reduction, evaluating their potential efficacy and relevance. We explore four VA RecSys engines that have shown superiority in uncovering complex semantic relationships of artwork and have been successfully applied for personalised recommendation tasks~\cite{Yilmaleiva23, YilmaleivaUMAP23}. In particular, we trained three uni-modal engines on image and textual data of artworks, and one multimodal engine that fuses both image and text. For our image-based approach we use the popular Residual Neural Network (ResNet)~\cite{he2016deep}, for our text-based approach we adopt both Latent Dirichlet Allocation (LDA)~\cite{blei2003latent} and Bidirectional Encoder Representations from Transformers (BERT)~\cite{devlin2018bert}, whereas for our fusion approach we use Bootstrapping Language-Image Pre-training (BLIP)~\cite{li2022blip}. The learned representations are used to derive personalised artwork recommendations for therapy that are presumably matching penitents' emotional needs. 
 
In sum, this paper makes the following contributions:
\begin{itemize}
    \item We develop and study four advanced VA RecSys engines using different backbone architectures 
    (ResNet, LDA, BERT, and BLIP)  to support PICS prevention and reduction through guided art therapy.

    \item We conduct a usability test with 4 healthcare experts to assess the appropriateness of VA RecSys engines 
    and a large-scale study with 150 post-ICU patients to assess the efficacy of the proposed VA RecSys engines,
    as compared to expert-curated recommendations.

    \item We contextualise our findings and provide guidance about potential strategies 
    to integrate ML-based VA RecSys in a personalised PICS intervention and beyond.
\end{itemize}

\section{Related work}
\label{sec:related-work}

\subsection{Approaches to prevent and reduce PICS}
There are several interventions focused on preventing PICS. The ABCDEF bundle~\cite{pun2019caring} is a commonly used strategy for preventing PICS which provides practical ways to promote patients’ state of being awake, autonomy, and rehabilitation. The bundle has demonstrated its effectiveness in improving the likelihood of survival, mitigating delirium, and reducing physical restraints in ICU patients~\cite{pun2019caring}. 

Environment management is another prevention strategy focusing on addressing environmental factors that are hostile to ICU patients which are associated with delirium and sleep deprivation, leading to PICS~\cite{inoue2019post}. Environment management includes the reduction of negative stimuli, such as overloaded background noise and lighting, which have been shown to improve sleep quality~\cite{demoule2017impact}\cite{litton2016efficacy} and reduce delirium among ICU patients~\cite{van2017effect}. The provision of positive stimuli is another way of environment management. Exposure to nature view, for example, demonstrated to shorten the length of hospital stay and reduce complications among patients~\cite{ulrich1984view}. In addition, music intervention reduces anxiety and stress-related measures in ICU patients~\cite{lee2017effects}. 

ICU diary is also a popularly practiced intervention supporting patients to understand what has happened during their ICU stay including times during sedation, which is effective in reducing anxiety, depression, and PTSD symptoms~\cite{inoue2019post}. Nevertheless, implementing all the aforementioned interventions requires systemic changes. Furthermore, due to their non-pharmacological nature, these prevention strategies often confront constraints related to increased workload for healthcare professionals~\cite{kotfis2020covid, ramnarain2021post}. 

Next to preventing PICS, it is also important to recognize early symptoms and provide interventions to avoid further development into, for example, acute stress disorder (ASD) and post-traumatic stress disorder (PTSD). ICU follow-up care is one of the most practiced interventions that address the psychological aspects of PICS~\cite{ramnarain2021post}\cite{inoue2019post}. ICU follow-up care aims to support patients through their transition period from hospital to home with guidance. This approach has gained significant traction in both Europe and North America in recent years and found to mitigate symptoms of PTSD stemming from ICU stay~\cite{jensen2015impact}. However, the current ICU follow-up care lacks a standardized structure, resulting in variations in interventions between hospitals, which makes it difficult to track the effectiveness of these interventions~\cite{jensen2015impact}. On the other hand, this variability paves the way for the advancement and implementation of individualized interventions, which often adopt technology. Overall, these interventions include guided meditation in natural settings using VR which could reduce pain and stress~\cite{pourmand2018virtual}, and gamified cognitive behavioral therapies for reducing depression~\cite{li2014game}. Furthermore, there exist various strategies to apply technology for ICU delirium prevention~\cite{kim2021overview}, many of which can be readily adapted to a post-intensive care environment. As such, there is the potential to enhance the rehabilitation of ICU patients through ICU follow-up care with a personalized methodology enabled by technology.

\subsection{Therapeutic visual experience and personalisation}
Since the study by~\citet{ulrich1984view}, which demonstrated the impact of having access to a nature view on enhancing the health outcomes of ICU patients, exposure to nature, and particularly its visual experiences, has gained interest as a nonpharmacological intervention~\cite{hamilton2010design}. Interventions are introduced wherein visual nature experiences play a role as a positive distraction~\cite{veling2021virtual,waszynski2018using}, generating positive feelings and maintaining attention for patients without inducing stress. 

Next to the exposure to nature, visual art is another popularly used positive distraction in hospital settings. A growing body of evidence suggested the impact of appropriate art on the health outcomes of patients \cite{hathorn2008guide}. Research also found that visual art with nature content is effective in reducing stress, anxiety, and perceived pain in critically ill patients \cite{ulrich2003healingarts,nanda2008undertaking}. Importantly, the characteristics of visual art, including subject matter and style, have been found to strongly influence its impact on patients \cite{ulrich2003healingarts,hamilton2010design}. Representative pictures dominated by nature, such as landscapes with trees and water, have been shown to be more effective in reducing anxiety and pain than other types, such as art without nature content \cite{ulrich1993effects}. Notably, the study showed that abstract art with rectilinear and straight-edge forms brings strong negative reactions \cite{ulrich2003healingarts}. 

The therapeutic effects of visual nature experiences are explained by evolutionary theory \cite{appleton1975experience} and the Biophilia hypothesis \cite{wilson1986biophilia}. These theories \cite{wilson1986biophilia,appleton1975experience}  propose that over millions of years of evolution, humans have become genetically predisposed to respond positively to natural settings that promoted well-being and survival for early humans. Negative effects of certain visual arts, on the other hand, can be explained by the Emotional Congruence theory \cite{ulrich2003healingarts}, which suggests that our perception of stimuli is influenced by our emotional states. This implies that abstract art, being open to interpretation, can lead patients in highly stressful situations to project negative emotions onto their interpretation. Consequently, they end up experiencing adverse visual experiences, as was demonstrated in Ulrich's study \cite{ulrich2003healingarts}, where immediate removal of such pictures became necessary. These instances underscore that while the use of art in a therapeutic setting is compelling, the application of art requires careful customization and personalization to ensure effectiveness. Recognizing these needs, Hathron \cite{hathorn2008guide}  emphasized the importance of considering individualized needs when selecting art for healing.

Art has been utilized outside of hospital settings, such as in the treatment of PTSD. Art therapy in these contexts takes various forms, ranging from exposure to creative techniques, with debatable effects \cite{schouten2015effectiveness}. In this study, we focus on the aspects of visual arts as a positive distraction. Taking this approach, we aim to improve temporary stress and evoke positive emotions among post-ICU patients, both during their time in the post-ICU clinic and throughout the rehabilitation stage. To enhance the positive impact, we incorporate techniques from narrative therapy \cite{madigan2011narrative} which encourages patients to explore deeper into the meanings of the visual arts and engage in prolonged exposure to the therapeutic elements of the art. We will further elaborate our approach in the method section.

\subsection{Visual art recommendation systems}
\label{subsec:va_lit}
VA RecSys represent an emerging field at the intersection of technology, art, and user preferences. These systems harness data-driven methodologies, particularly powered by machine learning algorithms, to facilitate personalised art recommendations, enabling users to discover artworks that resonate with their personal aesthetic inclinations. VA specifically paintings are both high dimensional and semantically complex, leading to a diverse and subjective set of emotional and cognitive reflections on users~\cite{wu2020paintkg}. Paintings reflect the complex and intricate interplay of concepts ranging from individual ideas and beliefs to concepts with cultural and historical significance of a society~\cite{palumbo2023visual}.  

With the proliferation of artworks in recent years, coupled with an increasing demand for personalization, VA RecSys has gained traction across diverse domains spanning from online platforms, to cultural heritage institutions aiming to enhance visitors experience and engagement. Studies like ~\cite{esman2012world, helal2013lessons} have shown a number of advantages and highlighted the significant role of VA RecSys. As the study of Falk et al.~\cite{falk2016identity} discuses, the main motivation of museum visitors is to have fun, experience art, learn new things, feel inspired, and interact with others. Thus, using VA RecSys empowered digital museum guides, visitors' expectations are not only to be exposed to artwork that matches their interest but also learn more and have access to more information~\cite{helal2013lessons}. Achieving this fundamentally requires VA RecSys to uncover complex semantics and abstract relationships between art works to derive recommendations. The earliest works such as Kuflik et al.~\cite{kuflik2014graph} and Deladiennee et al.~\cite{8022674} proposed a graph-based semantic VA RecSys
that relies on an ontological formalisation of knowledge. 

In recent years with the advancements in Artificial intelligence (AI) particularly Neural Networks (NN) demonstrating enormous success in capturing latent semantic structures and relationships from data, the RecSys domain started to adopt representation learning techniques~\cite{he2016vista}. Following this, a number of VA RecSys works emerged by learning latent representations from images of artworks ~\cite{messina2017exploring, messina2019content, messina2020curatornet} and from textual descriptions of paintings ~\cite{10.1145/3450613.3456847, yilma2020personalised} as well as from their combinations ~\cite{YilmaleivaUMAP23, Yilmaleiva23} demonstrating the power of NNs to derive meaningful and personalised recommendations. 
However, the integration of VA RecSys in healthcare and specifically within the context of PICS rehabilitation remains an unexplored domain. This paper aims to bridge this gap by investigating the potential of VA RecSys in aiding the art therapy process for PICS patients. By comparing VA RecSys-generated recommendations with expert-curated ones, this study seeks to enrich the current understanding of the role of VA RecSys in personalised healthcare interventions.

\section{Background: Learning latent representations of visual art}
\label{sec: background}
Representation learning is a powerful computational concept that involves automatically uncovering the underlying structure within complex data ~\cite{bengio2013representation}. It is a process where an algorithm learns to convert raw data inputs into more compact, meaningful, and feature-rich representations. These representations capture essential patterns, relationships, and characteristics within the data, enabling more effective analysis, understanding, and utilization ~\cite{zhong2016overview}.

In the context of VA RecSys, representation learning plays a pivotal role in converting intricate visual elements into condensed yet informative forms. This process involves training algorithms on text and/or image modalities, often based on neural networks, to recognise and extract not only distinctive features present in paintings, such as colours, shapes, and textures but also complex concepts embodied within artworks such as the emotional and cognitive reflections they trigger which are not always observable to the naked eye ~\cite{Yilmaleiva23}. The learned representations encode these features in a way that captures the essence of the artwork's semantics as well as visual identity. This goes beyond mere pixel values, as the algorithms internalise the higher-level characteristics that make each piece of art unique ~\cite{YilmaleivaUMAP23}. 

\begin{figure*}[!h]
\centering
\includegraphics[width=\textwidth]{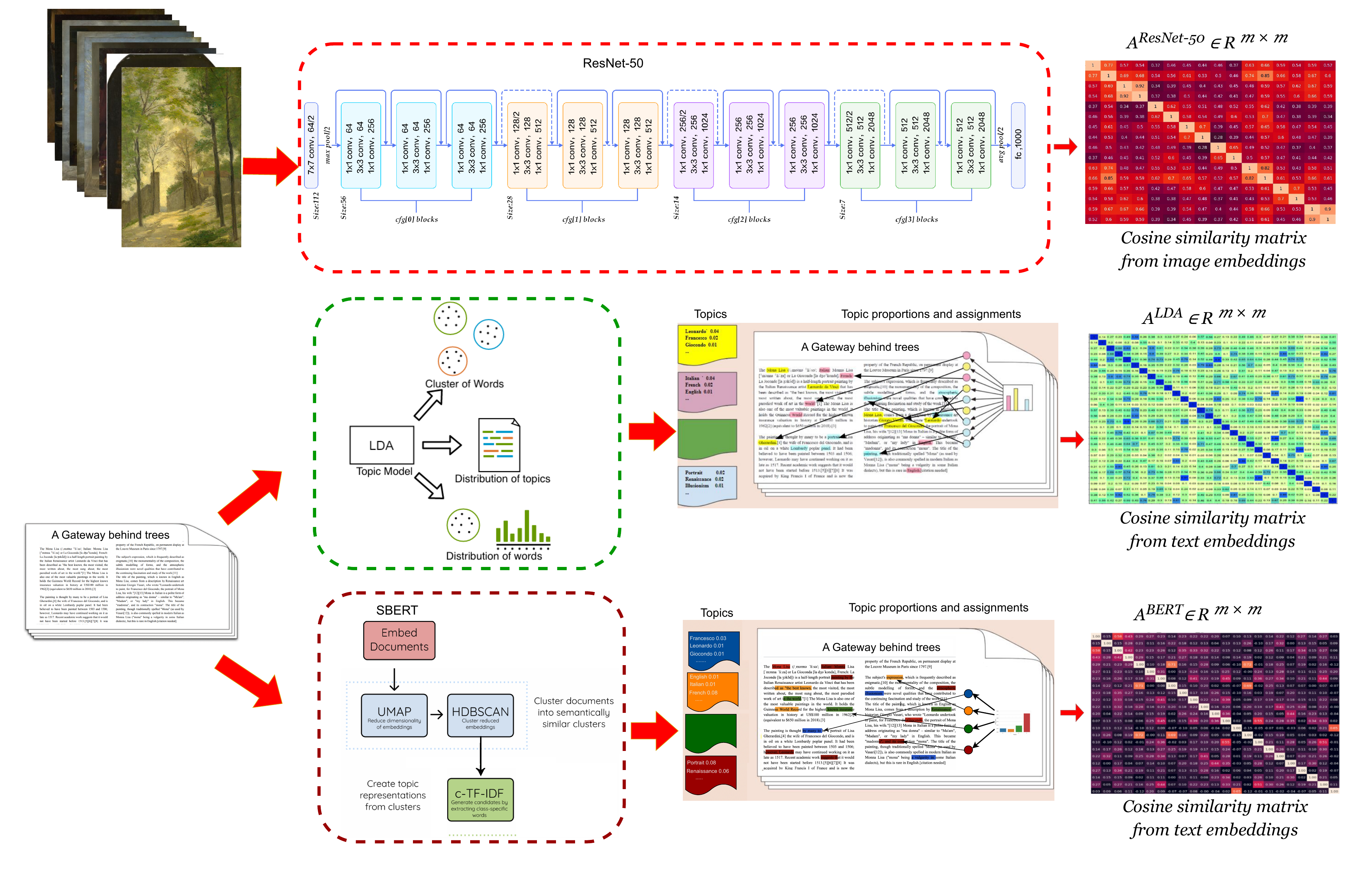}
\caption{Overview of our unimodal approaches to learning latent representations of paintings.}
\label{fig:feature_learning}
\end{figure*}

Based on the input data source there are two notable paths in representation learning literature which are unimodal and multimodal approaches~\cite{cooney2022unimodal}. As discussed in \autoref{subsec:va_lit}, we draw inspiration from the recent successes of VA RecSys employing NN-based representation learning, and observing how the resulting personalised recommendations capture hidden semantics and benefit users in many ways such as learning, discovery, enhanced engagement and better interaction experience. The key idea of representation learning in this setting is that textual and visual modalities of paintings are used to learn an embedding space where similar items are represented close to each other in the embedding space as explained in the following subsections. \autoref{fig:feature_learning} and \ref{fig:blip_feature_learning} summarise the painting representation learning approaches we propose and study for PICS rehabilitation therapy. 


\subsection{Unimodal VA representation learning}
This approach extracts and encodes inherent features of paintings from a single type of data modality (i.e., image or textual description).

\subsubsection{Image-based VA representation learning}
\label{subsub:Resnet}

Today, image feature extraction techniques predominantly rely on pre-trained Convolutional Neural Network (CNN) architectures, such as AlexNet~\cite{krizhevsky2017imagenet}, GoogLeNet~\cite{szegedy2015going}, and VGG~\cite{simonyan2014very}. An exemplar of this trend is the winner of the 2015 ImageNet challenge, ResNet, introduced by He et al.\cite{he2016deep}. ResNet pioneered the integration of residual layers to facilitate the training of very deep CNNs, setting the record with architectures comprising over 100 layers. A prominent version of this architecture is ResNet-50, featuring 50 layers, trained extensively on a vast repository of images from the ImageNet database.\footnote{\url{https://www.image-net.org}} Consequently, ResNet-50 has assimilated intricate feature representations across a diverse spectrum of images and has showcased its potential as a superior visual feature extractor compared to other pre-trained models\cite{ikechukwu2021resnet,li2021facial,barata2018survey}.

To extract latent visual features (image embeddings) from paintings, we employed the ResNet-50 model pre-trained on the ImageNet dataset. By channelling each painting image through the network, we derived a convolutional feature map, resulting in a feature vector representation. Upon completing the extraction process for all image features within the dataset containing $m$ number of images, we produce a matrix $\mathbf{A} \in \mathbb{R}^{m \times m}$, with each entry reflecting the cosine similarity measure among all image embeddings.  Cosine similarity is an effective metric to find item similarities from embedding spaces, which is commonly used in data mining and information retrieval~\cite{nguyen2019advanced, YilmaleivaUMAP23}. This matrix encapsulates the latent visual distribution across all images, serving as a foundation for calculating similarities among paintings for a VA RecSys task discussed in \autoref{sec: method}

\subsubsection{Text-based VA representation learning}

Learning latent representations of paintings from their textual descriptions was proven a powerful technique to uncover hidden semantic concepts that are embodied across artwork~\cite{10.1007/978-3-319-67162-8_35, yilma2020personalised, Yilmaleiva23}. In this work, we adopt two of the popular text-based representation learning approaches that demonstrated success in VA RecSys tasks namely, Latent Dirichlet Allocation (LDA) and Bidirectional Encoder Representations from Transformers (BERT). 

\paragraph{LDA} Our first text-based VA RecSys approach is LDA, an unsupervised generative probabilistic model proposed by Blei et al.~\cite{blei2003latent}. LDA attempts to model a collection of observations as a composite of distinct categories, or topics. In this context, each observation corresponds to a document, and the features are the presence, occurrence, or count of words, while the categories constitute the underlying topics. Notably, the specifics of the topics are not predefined; only the number of topics is chosen beforehand. These topics are learned as probability distributions over the words within each document. 

The procedure for constructing an LDA model within the VA RecSys framework is as follows. We begin by curating a collection of documents, each containing textual information about individual paintings. Subsequently, a desired number of topics, denoted as $k$, is determined, and each word $w$ within the document collection is assigned to a topic. This assignment is guided by $\theta_{i} \sim Dir(\alpha)$, where $\theta$ signifies the topic distribution for a document $d$, $\alpha$ represents the per-document topic distribution, $i \in {1, ..., k}$, and $Dir(\alpha)$ denotes a Dirichlet distribution spanning the $k$ topics. The learning phase involves computing conditional probabilities $P(t|d)$ (representing the likelihood of topic $t$ given document $d$) and $P(w|t)$ (indicating the likelihood of word $w$ given topic $t$). A comprehensive discourse on LDA topic modeling is presented in~\cite{blei2003latent} and~\cite{jelodar2019latent}.  Upon completing the training of the LDA model over the entire textual dataset containing $m$ number of documents representing each painting, a matrix $\mathbf{A} \in \mathbb{R}^{m \times m}$ is generated. Each entry $a(i,j)$ within this matrix corresponds to the cosine similarity measure between document embeddings. This matrix encapsulates the latent distribution of topics across all documents, which is utilised for calculating semantic similarities among paintings to derive recommendations.

\paragraph{BERT} Similarly, for the second approach with BERT, we start by curating documents for each painting. Then the feature learning process goes through three distinct phases. Firstly, we transform each painting document into an embedding representation by leveraging the pre-trained SBERT large language model.\footnote{In our implementation, we employed the \texttt{all-MiniLM-L6-v2} version to optimize performance, though alternative versions can also yield suitable painting embeddings.} This transformation maps sentences and paragraphs into a 384-dimensional dense vector space~\cite{DBLP:journals/corr/abs-1908-10084}. Secondly, we employ the uniform manifold approximation and projection (UMAP) algorithm~\citep{mcinnes2018umap} to reduce the dimensionality of these embeddings. UMAP, a dimension reduction technique, facilitates the transformation of multi-dimensional data points into a two-dimensional space. This step enhances efficiency while preserving the original embeddings' overarching structure. Thirdly, we leverage the HDBSCAN algorithm~\cite{campello2013density}, a soft-clustering technique, to semantically cluster the reduced embeddings. HDBSCAN avoids the misallocation of unrelated documents to clusters, thus enhancing the quality of clustering outcomes. 

From these clusters, we extract latent topic representations using a custom class-based term frequency-inverse document frequency (c-TF-IDF) algorithm. This algorithm generates importance scores for words within a topic cluster. The essence of c-TF-IDF lies in its capacity to provide topic descriptions by identifying the most vital words within a cluster. Words boasting high c-TF-IDF scores are selected for each topic, thereby creating topic-word distributions for every document cluster. A more detailed discussion of our topic modeling strategy with BERT can be found in the work of Grootendorst et al.~\cite{grootendorst2022bertopic}. Similar to the LDA approach, upon the completion of training the BERT model across the entire textual dataset of size $m$, we produce a matrix $\mathbf{A} \in \mathbb{R}^{m \times m}$. Each entry within this matrix quantifies the cosine similarity measure between all document embeddings. As with the LDA approach mentioned above, this similarity matrix captures the latent distribution of topics throughout all documents. Thus, it can be utilised to compute similarities of paintings for a recommendation task. 

\subsection{Multimodal VA representation learning}

This approach combines information from multiple data sources, like images and associated textual descriptions to create a unified representation space~\cite{Ramachandram17}. This joint embedding enables the exploration of the interconnectedness between the inherent attributes of each modality. The latent features extracted from images and textual descriptions are mapped into the same embedding space, ensuring that semantically similar images and corresponding textual descriptions are brought closer together~\cite{li2021align}. This synergy between textual narratives and visual aesthetics enhances the potential for various applications in interpreting artworks. Among the different approaches in the literature, we use Bootstrapping Language-Image Pre-training (BLIP)~\cite{li2022blip}, which has demonstrated superior performance in various downstream tasks, including VA RecSys.

BLIP is a technique that trains neural networks by combining language and image data. It trains a model to predict either an image or text given the other, in order to improve the model's understanding of multimodal relationships. BLIP uses a unified encoder-decoder model that can operate in three modes. The first mode, the unimodal encoder, encodes image and text separately. The second mode, the image-grounded text encoder, uses cross-attention to inject visual information into the text encoder. The third mode, the image-grounded text decoder, replaces bi-directional self-attention layers with causal self-attention layers. During pre-training BLIP optimizes three objectives: Image-Text Contrastive Loss (ITC), Image-Text Matching Loss (ITM), and Language Modeling Loss (LM). ITC aligns the visual and text transformers by encouraging similar representations for positive image-text pairs and dissimilar representations for negative pairs. ITM classifies whether image-text pairs are positive or negative. LM generates textual descriptions based on images. 

\begin{figure*}[!h]
\centering
\includegraphics[width=0.9\textwidth]{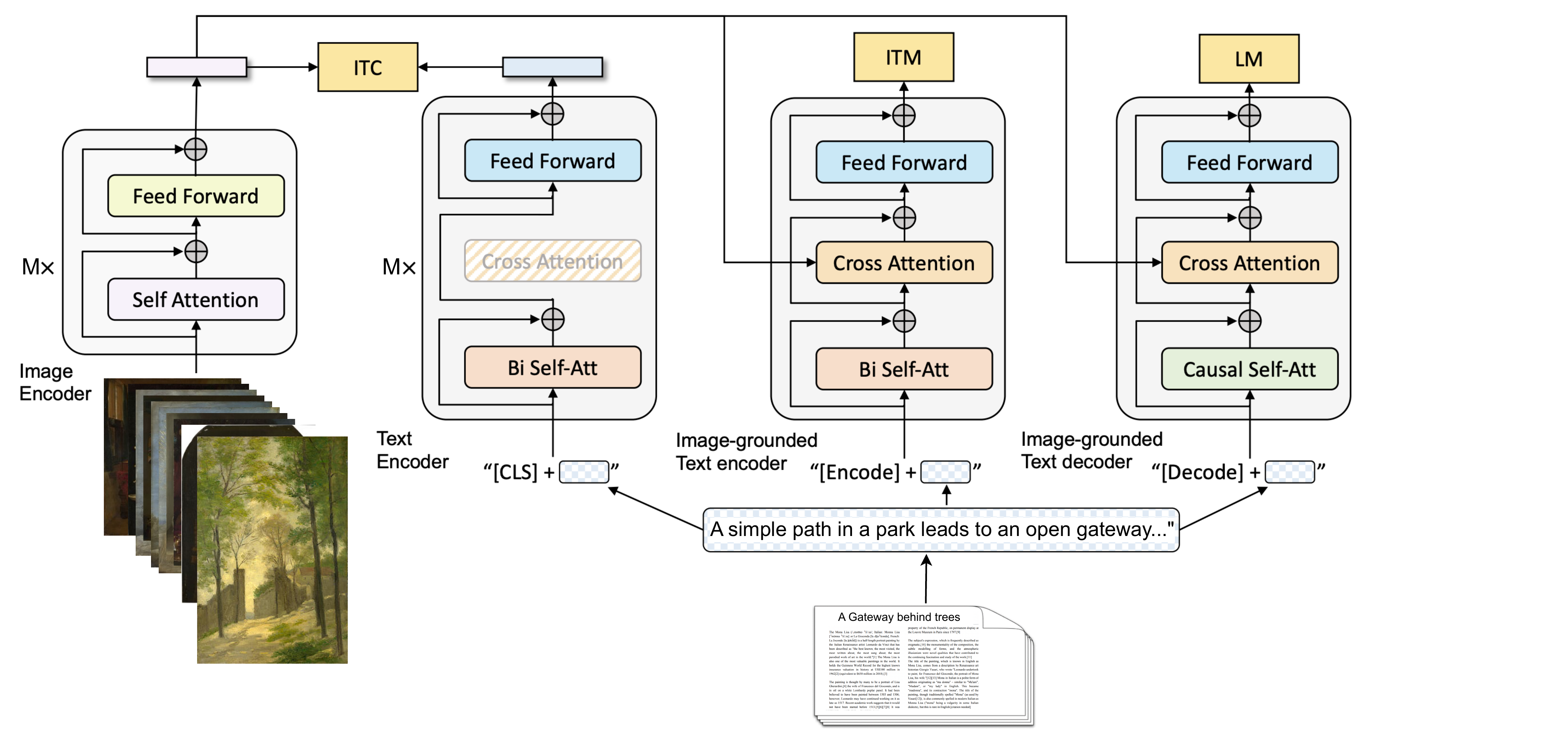}
\caption{Overview of our multimodal approach with BLIP to learning latent semantic representations of paintings.}
\label{fig:blip_feature_learning}
\end{figure*}

For our VA representation learning task, we utilized the pre-trained BLIP model as a multimodal feature extractor. First, we extract multimodal features and use the ITM head to compute ITM scores for each painting, generating probability-matching scores for each image-text pair. Then, we compute a matrix $\mathbf{A} \in \mathbb{R}^{m \times m}$ where each entry $\mathbf{A}_{ij}$ is the probability matching score between the joint painting embeddings which can be used to compute similarities for a VA RecSys tasks. See \autoref{fig:blip_feature_learning} for an illustration of our multimodal approach to learning latent semantic representations of paintings with BLIP. 

\section{Method: Personalised visual art recommendation for PICS therapy}
\label{sec: method}

To enhance the potential therapeutic benefits of a visual art experience, we designed a personalized guided art therapy. To tap into participants' latent needs and preferences, the process starts with inviting them to choose their preferred painting from sample paintings. Based on their selections, participants were subsequently presented with a set of three paintings, carefully chosen to align with their preferred sample painting. This inclusion of multiple paintings was intended to extend the duration of exposure to the painting. Additionally, for each of these selected paintings, we provided accompanying text guidance to facilitate active engagement and thoughtful reflection among participants. In this study, we explore two different approaches of personalised VA recommendation strategies for PICS art therapy; Expert-based and VA RecSys-based recommendation.  

\subsection{Expert recommendations}

This established approach involves the curation of recommendations by experienced clinicians. Their expertise allows them to choose artworks that elicit desired emotional and psychological responses, in line with PICS rehabilitation therapy objectives. The recommendation procedure is primarily guided by the patients' preference among a list of alternative paintings presented to them in order to identify the paintings that they resonate most profoundly with their journey towards recovery.  Following this clinicians will select a set of recommended paintings that evoke similar emotions and moods that align with the therapeutic goals of the recovery journey.

\subsection{VA RecSys recommendations}
We consider unimodal and multimodal representation learning techniques discussed in \autoref{sec: background} that can learn features from both textual and visual modalities paintings. Particularly we study four models; LDA and BERT to learn text-based representations, ResNet for image-based representations, and BLIP for the fusion of the two modalities.  

\begin{figure*}
\includegraphics[width=\textwidth]{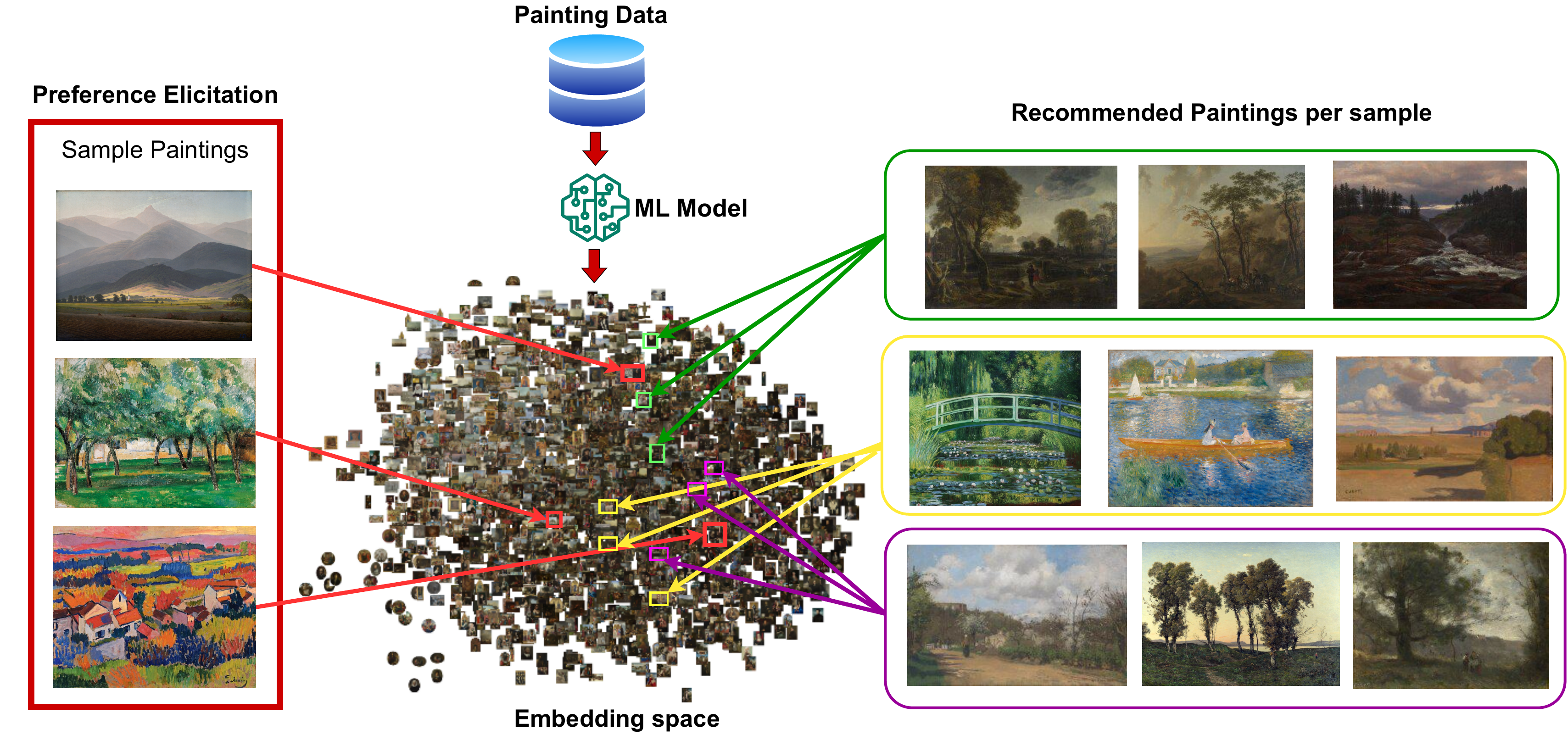}
\caption{Overview of our proposed ML-based VA RecSys pipeline. Sample paintings to recall ICU experience (left), our ML-based VA RecSys engines (Middle) and healing painting recommendations (right) }
\label{fig: VA_RecSys_pipeline}
\end{figure*}

Let $P = \{p_1, p_2, \dots, p_m\}$ be a set of image paintings
and $\mathcal{P} = \{\vp_1, \vp_2, \dots, \vp_m\}$ be the associated embeddings of each painting according to LDA, BERT, ResNet or BLIP. 
Once the dataset embeddings (latent feature vectors) are learned using either model (LDA, BERT, ResNet or BLIP) we compute the similarity matrix for all the paintings $\mathbf{A}$. Next, the preference of a patient user $u$ is modelled by computing a ranking score for the paintings in the dataset according to their therapeutic relevance (i.e. similarity to the painting $p_j$ a user indicated to support their recovery). Thus, the predicted score $S^u(p_i)$ the user would give to each painting in the collection $P$
is calculated as: 
\begin{equation}\label{eq:user-score}
    S^u(p_i) = d(\vp_i, \vp_j)
\end{equation}

where $d(\vp_i, \vp_j)$ is the cosine similarity between embeddings of paintings $p_i$ and $p_j$ in the computed similarity matrix.   
Once the scoring procedure is complete, the paintings are sorted and the $r$ most similar paintings constitute a ranked recommendation list. 
\autoref{fig: VA_RecSys_pipeline} summarises our VA RecSys-based recommendation pipeline. 

\section{Materials}

\subsection{Sample paintings for preference elicitation}
To elicit user preferences, we offer sample paintings, allowing users to provide their input by choosing one of them. These sample paintings were derived from a pre-study approved by the Ethics Review Panel of the University of Twente, conducted with a total of 186 former patients, including 10 former ICU patients. 

In a pre-study, an initial selection of 18 nature paintings associated with positive emotions, such as relaxing, cheerful, and awe-inspiring, was made by an academic expert in affective design from WikiArt.\footnote{\url{https://www.wikiart.org}} These paintings varied in style including different levels of abstraction. Subsequently, this set of paintings was narrowed down to 6, each of which was found to strongly evoke one or multiple positive emotions. This selection process involved evaluation by three experts in affective psychology, environmental psychology, and affective and healthcare design, each with over 15 years of experience in their respective fields. 

Using these final six paintings, we conducted an online survey to ask participants to choose the painting that would best support their recovery. The results of the pre-study showed that Painting 2 was chosen by the majority of participants (n=63), followed by Painting 1 (n=32) and Painting 3 (n=32). These three paintings are shown in \autoref{fig: VA_RecSys_pipeline}, left column (labels as `sample paintings') form top to bottom. 
In the present study, we use these top three most frequently selected paintings as sample paintings for preference elicitation to derive personalised recommendations. These paintings encompass diverse styles and are believed to convey emotions associated with calmness, restoration, and cheerfulness, respectively, providing a comprehensive representation of the healing experience.

\subsection{Dataset for generating recommended paintings}
\begin{figure*}[!ht]
\centering
\includegraphics[width= 0.9\textwidth]{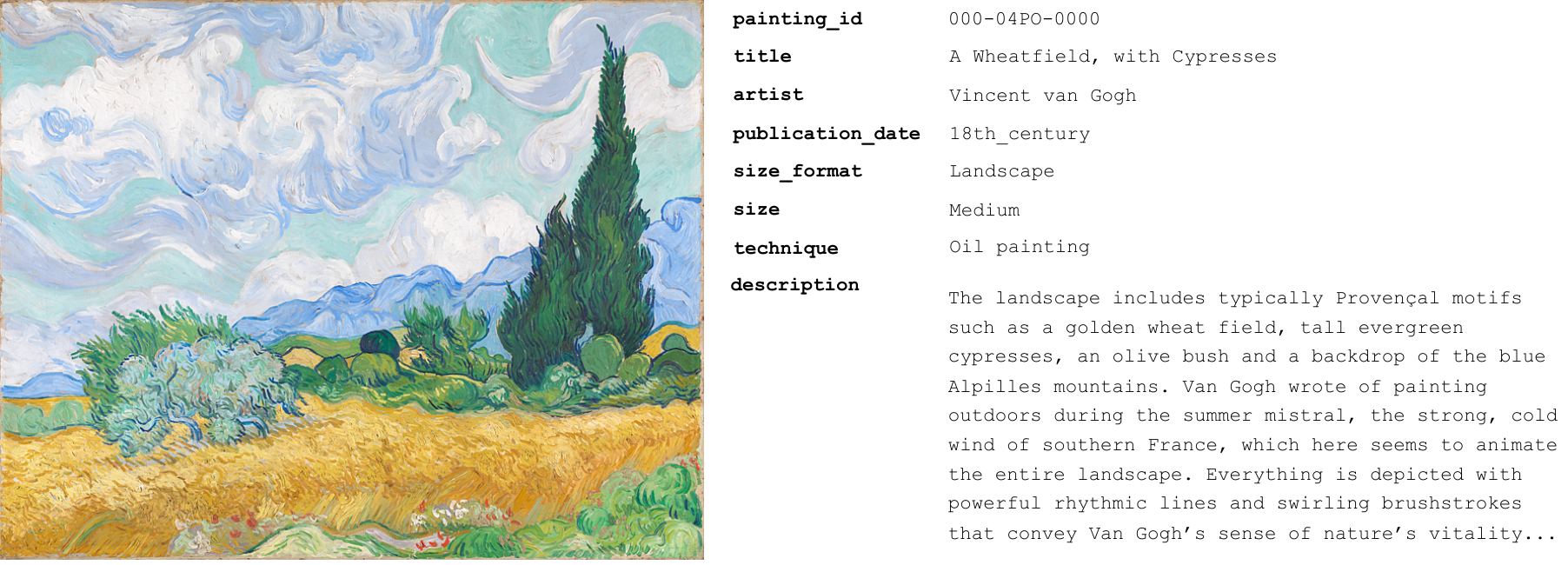}
\caption{Sample painting and associated metadata from the dataset.}
\label{fig:ng_sample}
\end{figure*}

We used a dataset combining the nature paintings from the pre-study with a collection containing 2,368 paintings from the National Gallery, London.\footnote{https://www.nationalgallery.org.uk/} provided through the CrossCult Knowledge Base.\footnote{\url{https://www.crosscult.lu/}} Every painting image within the dataset is accompanied by a complementary set of text-based metadata. This dataset's configuration renders it well-suited for evaluating the proposed feature learning methodologies. A representative data point is illustrated in Figure \ref{fig:ng_sample}.

For acquiring textual features via LDA and BERT models, we performed pre-processing of the painting metadata. This encompassed concatenating text fields, excluding punctuation symbols and stop-words, transforming to lowercase, and applying lemmatization. Conversely, for the acquisition of visual features through the ResNet model, we utilized the authentic painting images.\footnote{All paintings are available under a Creative Commons (CC) license.} This procedure involved extracting convolutional feature maps using the pre-trained ResNet-50 model. 
For the multimodal feature learning with BLIP, both pre-processed text and image data sources were jointly utilized, as elaborated in Section \ref{sec: background}.

\subsection{Ensuring a safe and sensitive deployment: Expert evaluation of VA RecSys engines}
\label{subsec: pilot}

Before delving into our comprehensive study involving end-users, particularly PICS patients, we conducted a pilot test with experts from diverse relevant domains, as a proactive measure to ensure their safety and well-being. The primary objective was to preempt any potential harm or inadvertent elicitation of negative emotions that might arise from the recommendations generated by the VA RecSys engines. The core purpose of this usability test was to ascertain the suitability of VA RecSys engine-generated recommendations for deployment without requiring human intervention. 

\subsubsection{Apparatus}
We created an overview of sample paintings and their respective recommendations generated by each of the engines. The recommendation engines were anonymized. Participants were provided one set of recommendations from each VA RecSys engine at a time. 

\subsubsection{Participants}
Four experts well-versed in diverse domains were recruited, including ICU nursing, healthcare design research, and affective design research.

\subsubsection{Design}
The experts were exposed to all recommendation pipelines (within-subjects) design and were asked to assess the appropriateness of each of the top-3 recommended paintings from all pipelines for all samples. 

\subsubsection{Procedure}
Participants were invited for a one-to-one interview to assess the VA RecSys engines recommendation. There, they were informed about the purpose of the study. Then participants were provided with individual recommendation paintings per sample. For each recommended painting, they were asked to respond to the question ``To what extent do the recommended painting align with the original painting in terms of the overall experience they evoke? This experience can be visual, semantic, or a combination of both.'' They responded in a 1-5 scale (1:\,not at all, 5:\,very much) as reported in \autoref{tab: pilot}. Furthermore, they were also asked to give their expert opinion on the appropriateness of the recommended paintings for the purpose of therapy. 

\subsubsection{Results}

By leveraging the expertise of recruited professionals, we meticulously assessed the engines to identify those that align harmoniously with therapeutic goals and emotional well-being. The insights garnered from this test served as a foundation for selecting engines that could be employed with a higher degree of confidence, thereby fostering an environment of safety and sensitivity in our subsequent studies involving PICS patients.

\begin{figure*}[!h]
\includegraphics[width=\textwidth]{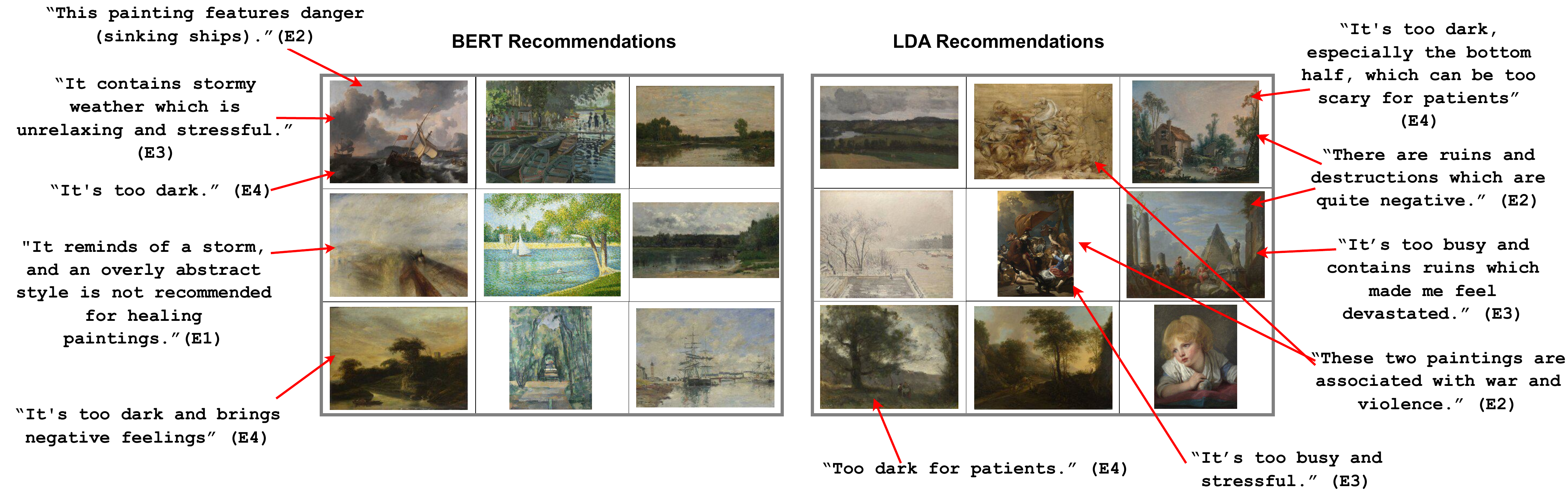}
\caption{Expert reflections on the top three direct recommendations from our text-based VA RecSys engines (BERT and LDA).}
\label{fig: pilot_ref}
\end{figure*}

\begin{figure*}[!h]
\includegraphics[width=\textwidth]{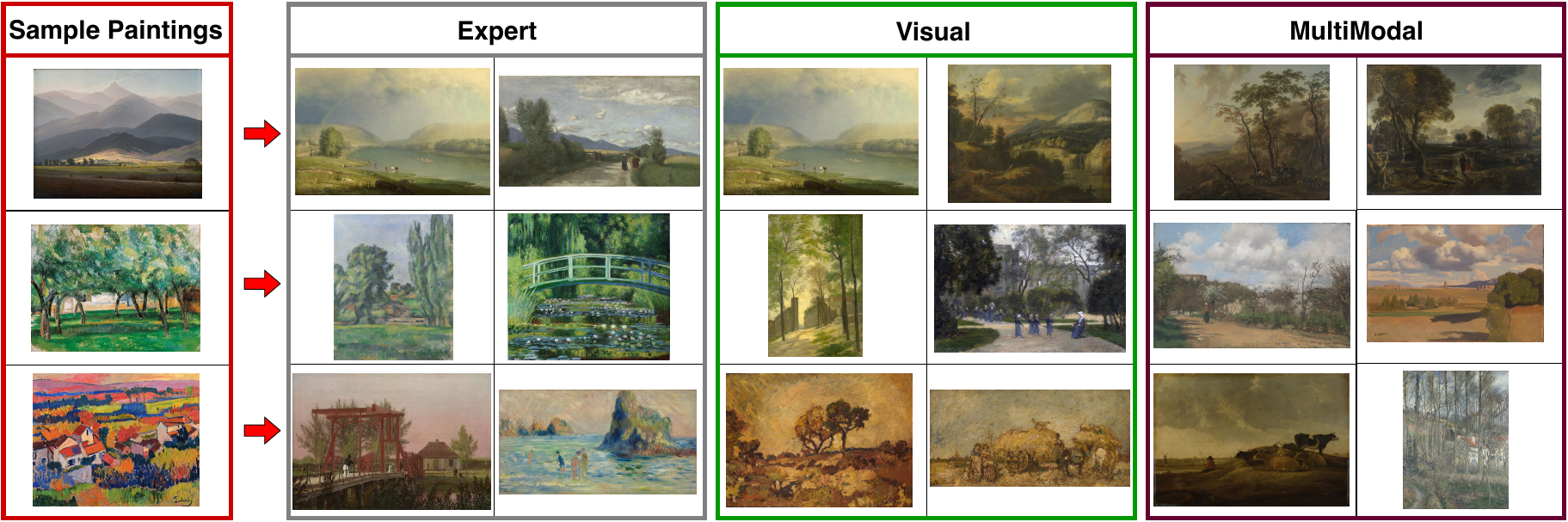}
\caption{Top-2 recommendations for each sample painting from expert-validated engines.}
\label{fig: sample_rec}
\end{figure*}

From the experts' evaluation, we observed that both text-based VA RecSys engines were not suitable for deployment in a real-world application. As they were found to be dark and evoking negative emotions. Some of the expert reflections on these recommendations are depicted in \autoref{fig: pilot_ref}. The figure shows some examples of the comments provided by the experts. As can be appreciated, some of the recommended images were too dark which could elicit fear or contain images of violence that could trigger traumatic experiences. Following this pilot we have decided to proceed with only three recommendation methods for our user study; Expert, Visual and Multimodal. \autoref{fig: sample_rec} shows examples of the top-2 recommendations from each of the expert validated approaches.

\begin{table*}[!h]
{%
\begin{tabular}{lcccccccccccc}
\toprule
& & \textbf{Fusion BLIP} & \textbf{Image ResNet} & \textbf{Text BERT} & \textbf{Text LDA} \\ 
\midrule
\multirow{3}{*}{\textbf{Expert 1 (Affective Design, ICU research, +10)}} & Painting 1 & 2 & 3 & 1 & 1 \\ 
& Painting 2 & 3 & 2 & 3 & 1 \\ 
& Painting 3 & 2 & 3 & 2 & 1 \\ 
\midrule
\multirow{3}{*}{\textbf{Expert 2 (Affective Design, +10)}} & Painting 1 & 3 & 2 & 2 & 1 \\ 
& Painting 2 & 3 & 2 & 1 & 2 \\ 
& Painting 3 & 2 & 4 & 2 & 1 \\ 
\midrule
\multirow{3}{*}{\textbf{Expert 3 (Affective Design, +10)}} & Painting 1 & 2 & 4 & 1 & 1 \\ 
& Painting 2 & 1 & 1 & 2 & 4 \\ 
& Painting 3 & 2 & 3 & 1 & 1 \\ 
\midrule
\multirow{3}{*}{\textbf{Expert 4 (ICU Nurse, +10)}} & Painting 1 & 1 & 4 & 2 & 2 \\ 
& Painting 2 & 4 & 2 & 3 & 3 \\ 
& Painting 3 & 2 & 3 & 2 & 1 \\ 
\midrule
\textbf{Total} & & \textbf{27} & \textbf{33} & \textbf{22} & \textbf{19} \\ 
\bottomrule \\
\end{tabular}%
}
\caption{In our pilot test with Experts, they were shown 3 paintings from each of the VA RecSys engines for all sample paintings.}
\label{tab: pilot}
\end{table*}

\section{Evaluation: User study}


The main goal of our evaluation was to understand the user's perception towards the quality of our studied VA recommendation strategies for PICS rehabilitation therapy, and ultimately to assess their efficacy in supporting the healing journey of PICS survivors. We conducted 
a large-scale user study, to be described later, that was approved by the Ethics Review Panel of the University of Twente. 

\subsection{Apparatus}
We designed an online guided therapy survey using Google Forms\footnote{\url{https://www.google.com/forms/about/}} that first elicited the preferences of participants by providing them with a set of paintings to choose from that resonates with their healing journey. Then participants were taken through a guided therapy session by using three paintings that were recommended based on the elicited preference choices.

\begin{figure*}[!ht]
    \centering

    \def\h{8cm}
    \includegraphics[height=\h]{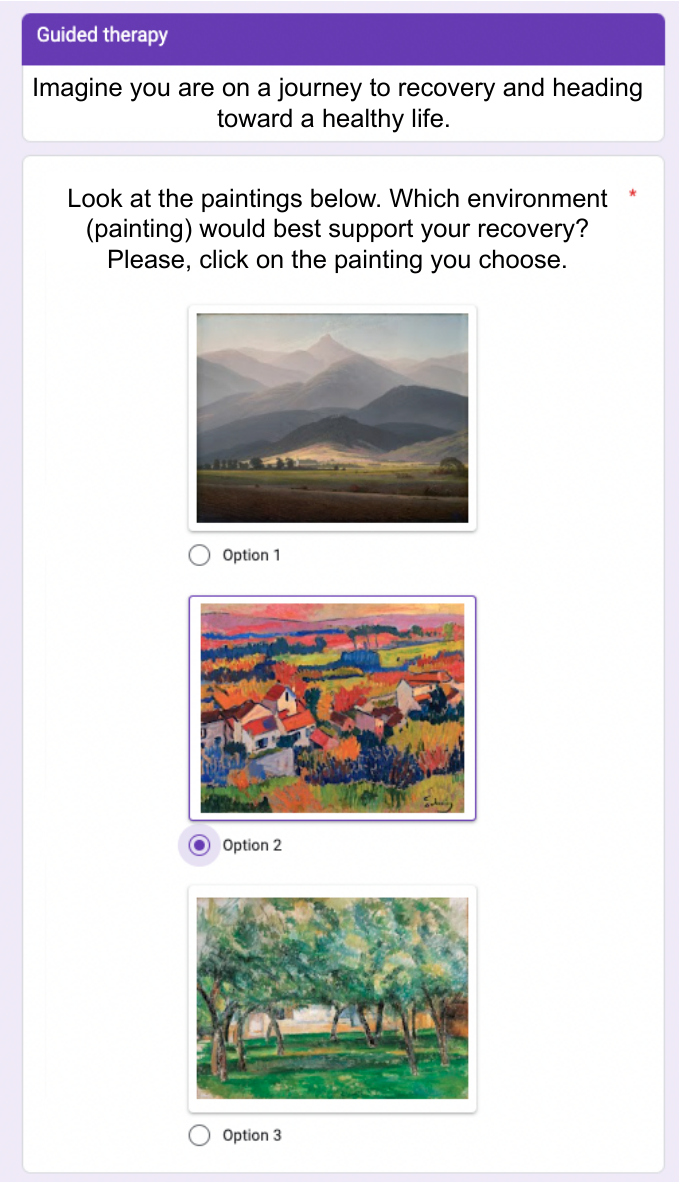} \hfil
    \includegraphics[height=\h]{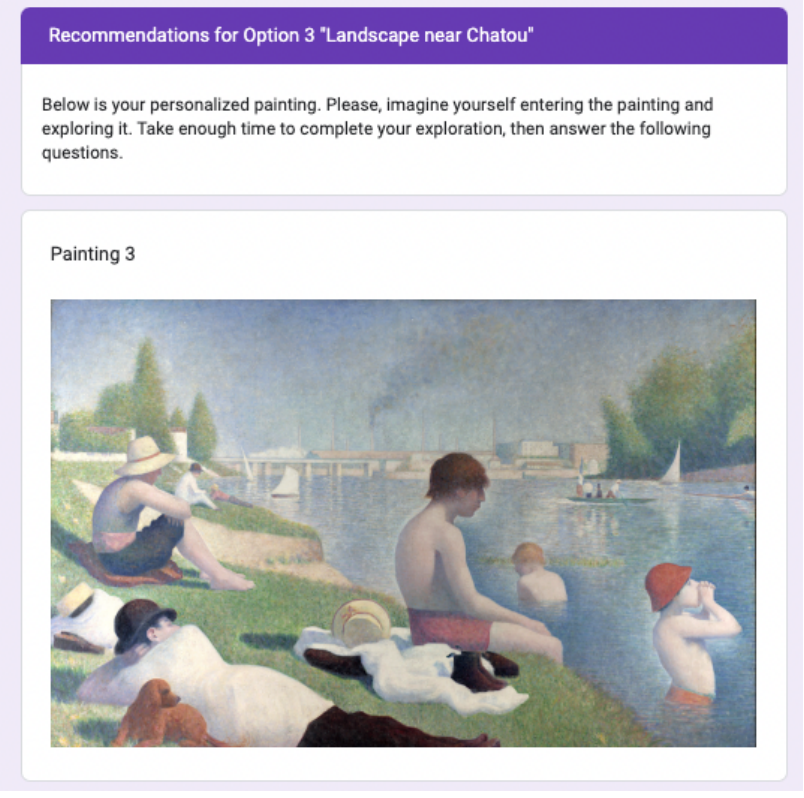}

    \caption{
        Screenshots of our online survey for guided therapy: preference elicitation (left) and recommendations (right).
    }
    \label{fig:webapp}
\end{figure*}
\subsection{Participants}

We recruited a representative pool of $N=150$ participants via the Prolific crowdsourcing platform.\footnote{\url{https://www.prolific.co/}} 
We identified people who had conditions in the past that may potentially lead to PICS, 
such as COVID-19 patients who were treated in a hospital, 
cancer survivors, 
or having other surgical issues. 
For this study, we focused on post-COVID patients, as the likelihood of having developed PICS recently is higher.
The main screening criteria was ``I have been officially diagnosed with COVID-19 (tested by a licensed medical professional), and was treated in a hospital.''.
Furthermore, we enforced other criteria for eligibility:  
being fluent in English, minimum approval rate of 100\% in previous crowdsourcing studies in the platform, 
and being active on the platform in the last 90 days.

Our recruited participants (74 female, 76 male) were aged 32.7 years (SD=10.9) and could complete the study only once. 
Most of them lived in UK (40 participants), USA (28), or South Africa (35).
Most participants had been through general ward or a medical/surgical unit (67), 
or had been in ICU (44) or in an Emergency Room (37).
For most participants, the duration of their stay in a hospital was 
less than a week (71) or between 1 to 2 weeks (50).
Most participants declared to suffer from anxiety (130) and/or depression (127), indicating the presence of psychological components related to PICS symptoms. 
The study took a median time of 23\,min to complete and participants were paid an equivalent hourly wage of \$12/h.  We also administered the Patient Health Questionnaire-4 (PHQ-4)~\cite{lowe20104} to collect signs of psychological symptoms related to PICS: anxiety and depression. Most participants (86\%) exhibited at least one symptom, with the majority (70\%) showing symptoms for both of these conditions.

\subsection{Design}

Following the expert evaluation of our developed VA RecSys engines discussed in \autoref{subsec: pilot}, 
we deployed two engines together with the expert recommendations: 
visual (ResNet-based VA RecSys) and multimodal (BLIP-based VA RecSys). 
Each participant was only exposed to the recommendations generated 
by one of the three groups (between-subjects design) and then went through guided art therapy. 
Each group comprised 50 participants.

\subsection{Procedure}
We first assessed baseline and post-test affective states using two different measures: a Pick-A-Mood tool \cite{desmet2016mood} for moods and the short version of the Positive and Negative Affect Schedule (PANAS) scale~\cite{watson1988development} for emotions. In order to reduce the cognitive load on participants and to be more efficient, we selected 10 items that are relevant to PICS (for negative emotions) and patient well-being (for positive emotions). Particularly we consider 5 positive and 5 negative items (instead of 10 positive and 10 negative items) together with a neutral item to assess the affective state of participants before and after guided art therapy. 

Guided art therapy involves asking questions that encourage participants to engage with the paintings, 
such as \textit{``Imagine yourself entering the painting and exploring it. How did you feel while spending time in this painting?''} 
Participants were prompted to reflect on their experience and describe it in three to four sentences. 

Participants rated the provided painting recommendations in a 5-point Likert scale. 
Our dependent variables are widely accepted proxies of recommendation quality~\cite{pu2011user}:
\begin{description}
    \item[Accuracy:] The paintings match my personal preferences and interests.
    \item[Diversity:] The paintings are diverse.
    \item[Novelty:] I discovered paintings I did not know before.
    \item[Serendipity:] I found surprisingly interesting paintings.
\end{description}

We also collected two dependent variables that inform to what extent the recommended paintings 
contributed to a sense of immersion and engagement:
\begin{description}
    \item[Immersion:] How much do the recommended paintings contribute to your sense of immersion, 
        making you feel deeply involved or absorbed in the artwork?
    \item[Engagement:] To what extent do the recommended paintings contribute to your feeling of engagement, 
        capturing your attention and generating a sense of involvement or interest?
\end{description}

\subsection{Results}
 
\subsubsection{Comparison of recommended groups: Expert, Visual, and Multimodal}
We investigated whether there were differences between the three recommendation groups (i.e. Expert, Visual, and Multimodal).
We used a linear mixed-effects (LME) model where each dependent variable is explained by each recommendation group.
Participants are considered random effects.
An LME model is appropriate here because the dependent variables are discrete and have a natural order.

We fitted the LME models (one model per dependent variable) 
and computed the estimated marginal means for specified factors.
We then ran pairwise comparisons (also known as \emph{contrasts})
with Bonferroni-Holm correction to guard against multiple comparisons.

\paragraph{Analysis of recommendation quality measures}

\begin{figure*}[!ht]
    \centering
    \def\w{0.24\linewidth}
    
    \subfloat[Immersion]{\includegraphics[width=\w]{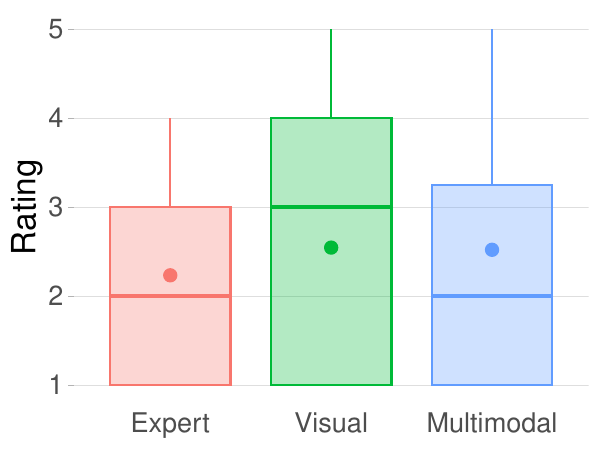}}\hfill
    \subfloat[Engagement]{\includegraphics[width=\w]{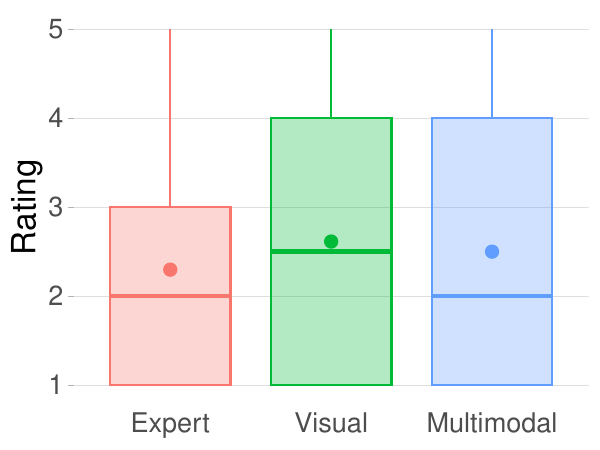}}\hfill
    \subfloat[Accuracy]{\includegraphics[width=\w]{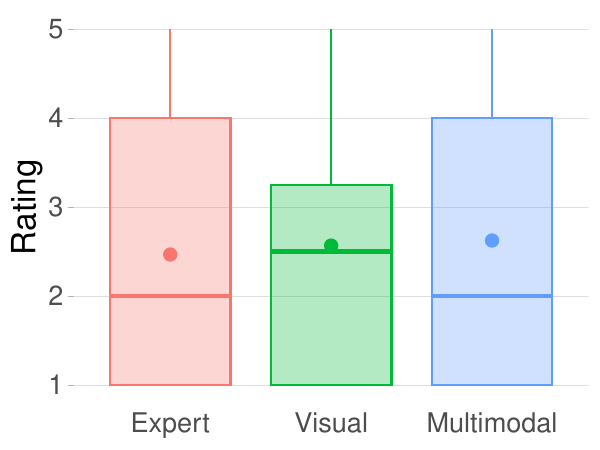}} \\
    \subfloat[Diversity]{\includegraphics[width=\w]{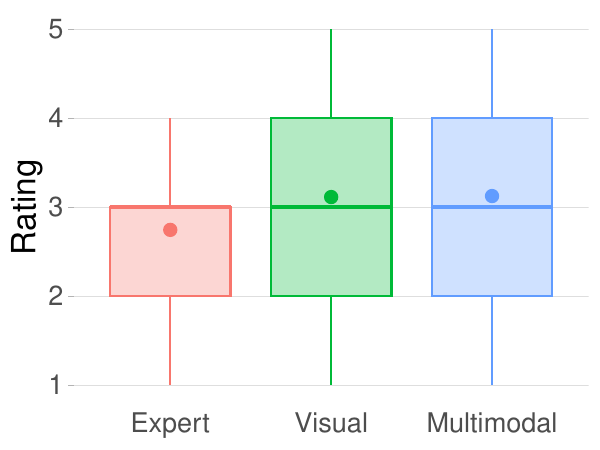}} \hfill
    \subfloat[Novelty]{\includegraphics[width=\w]{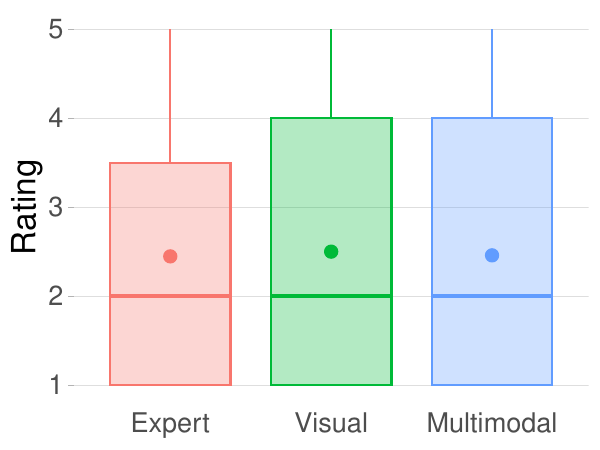}} \hfill
    \subfloat[Serendipity]{\includegraphics[width=\w]{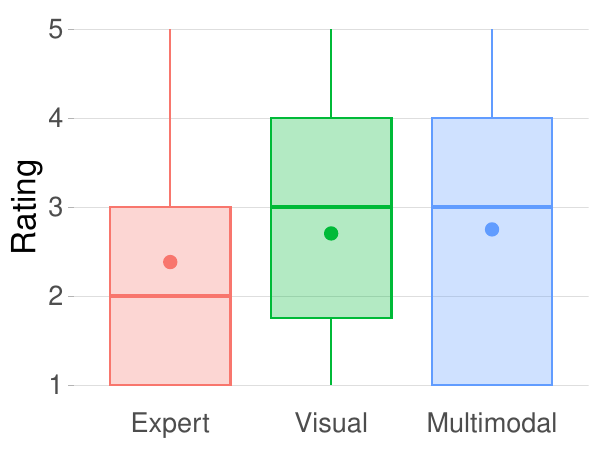}} 

    \caption{Distribution (box plots) of user ratings for the user-centric dependent variables of recommendation quality. Dots denote mean values.}
    \label{fig:prolific-ratings-overall}
\end{figure*}

\autoref{fig:prolific-ratings-overall} shows the distributions of user ratings
for the user-centric dependent variables of recommendation quality.
Differences between groups were not statistically significant in any case, with small to moderate effect sizes.
The largest effect sizes were observed when comparing Expert and Multimodal recommendations
in terms of diversity ($r=0.177$) and serendipity ($r=0.119$).

\paragraph{Analysis of changes in mood}

\begin{figure*}[!ht]
    \centering

    \def\w{0.8\textwidth}
    \includegraphics[width=\w]{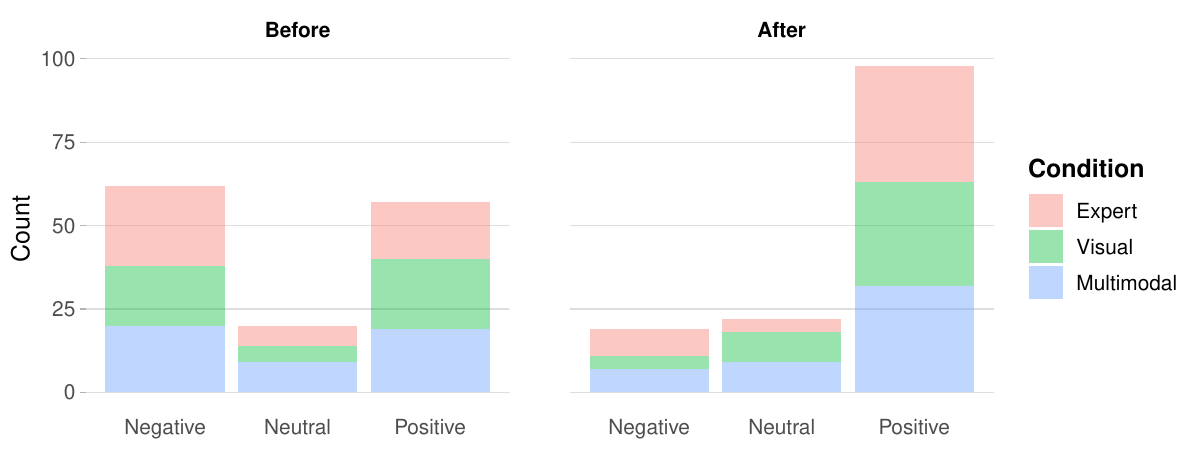}

    \caption{
        Mood improvement comparison before and after art therapy.
    }
    \label{fig:mood_improvement}
\end{figure*}

\autoref{fig:mood_improvement} shows the mood changes before and after therapy,
for the three recommendation groups we have considered in our study. In all three groups, a mood enhancement effect of guided art therapy was observed. When comparing it to the baseline where the majority of participants were in a negative mood (44.6\%), after guided art therapy the majority of participants reported being in a positive mood (70.5\%), with only a minority remaining in a negative (13.6\%) or neutral (15.8\%) mood.

\begin{figure*}[!ht]
    \centering
    \def\w{0.2\linewidth}

    \includegraphics[width=0.9\linewidth]{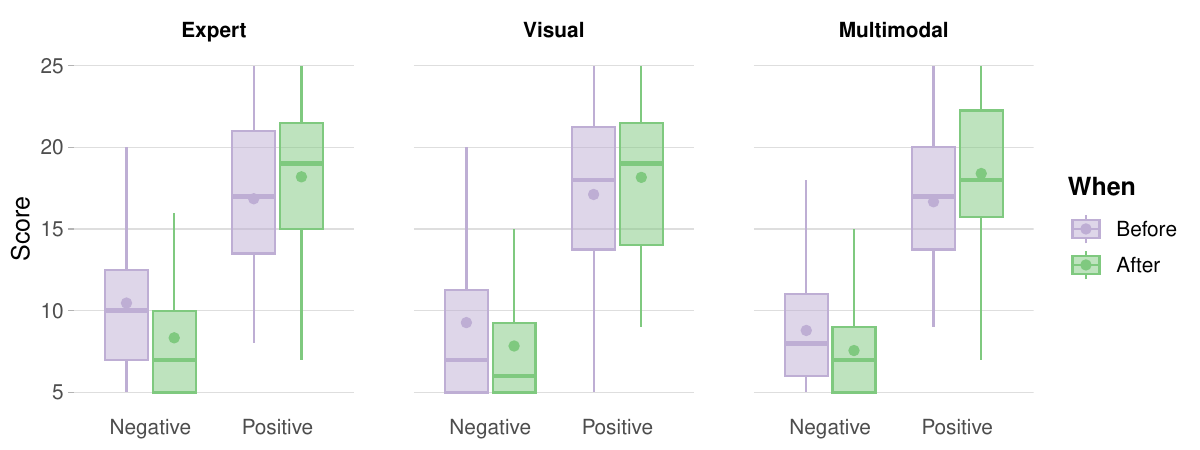}
    \caption{Emotion score changes according to the Positive Affect Negative Affect Schedule (PANAS) Scale.} 
    \label{fig:prolific-ratings-diff}
\end{figure*}

\autoref{fig:prolific-ratings-diff} shows the change in scores after therapy,
aggregated according to the ten items of the PANAS scale.
Differences between groups were not statistically significant in any case, with small to moderate effect sizes.
According to an item-independent analysis,
the largest effect sizes were observed in terms of the `afraid' item,
when comparing Expert recommendations against Visual ($r=0.181$) and Multimodal ($r=0.175$) recommendations,
followed by the `scared' item when comparing Expert and Visual recommendations ($r=0.122$). 
Upon further examination, users in the Visual group did not change
their scores for the `afraid' item (the median difference is 0).
This was also the case for users in both the Visual and Multimodal group
with regards to the `scared' item.

\subsubsection{Analysis of user reflections}

\begin{figure*}[!ht]
    \centering

    \def\w{\textwidth}
    \includegraphics[width=\w]{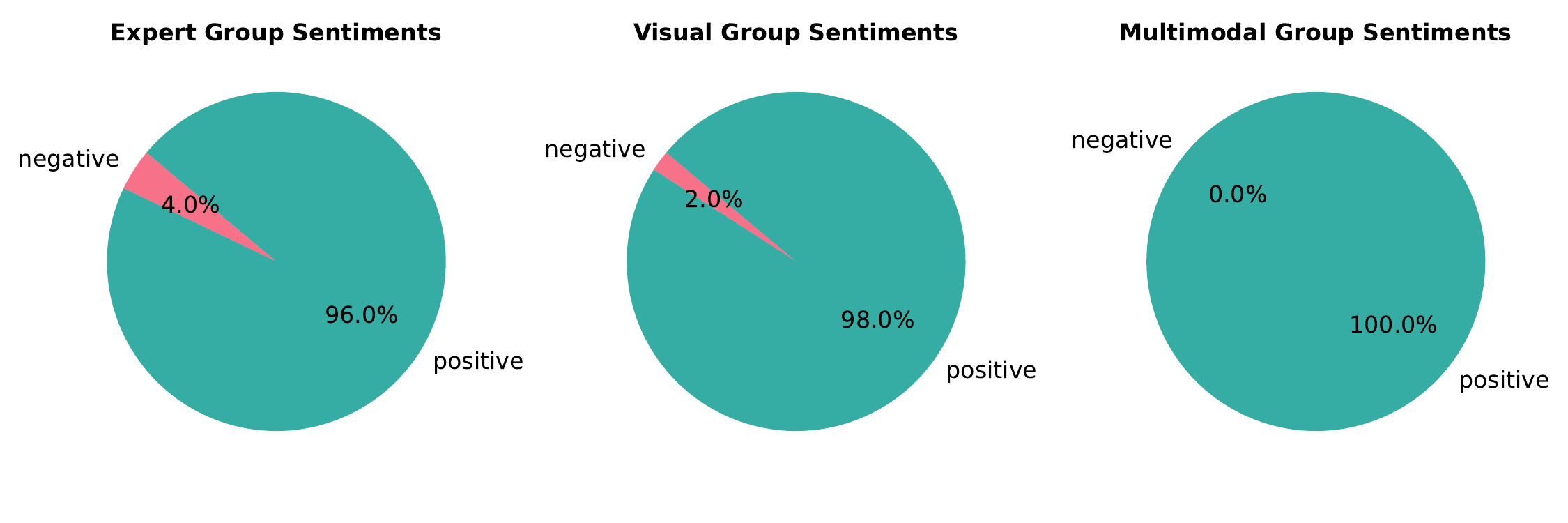}

    \caption{
        Sentiment analysis of user reflections per group.
    }
    \label{fig:sentiment}
\end{figure*}

\begin{figure*}[!ht]
    \centering

    \def\w{0.9\textwidth}
    \includegraphics[width=\w]{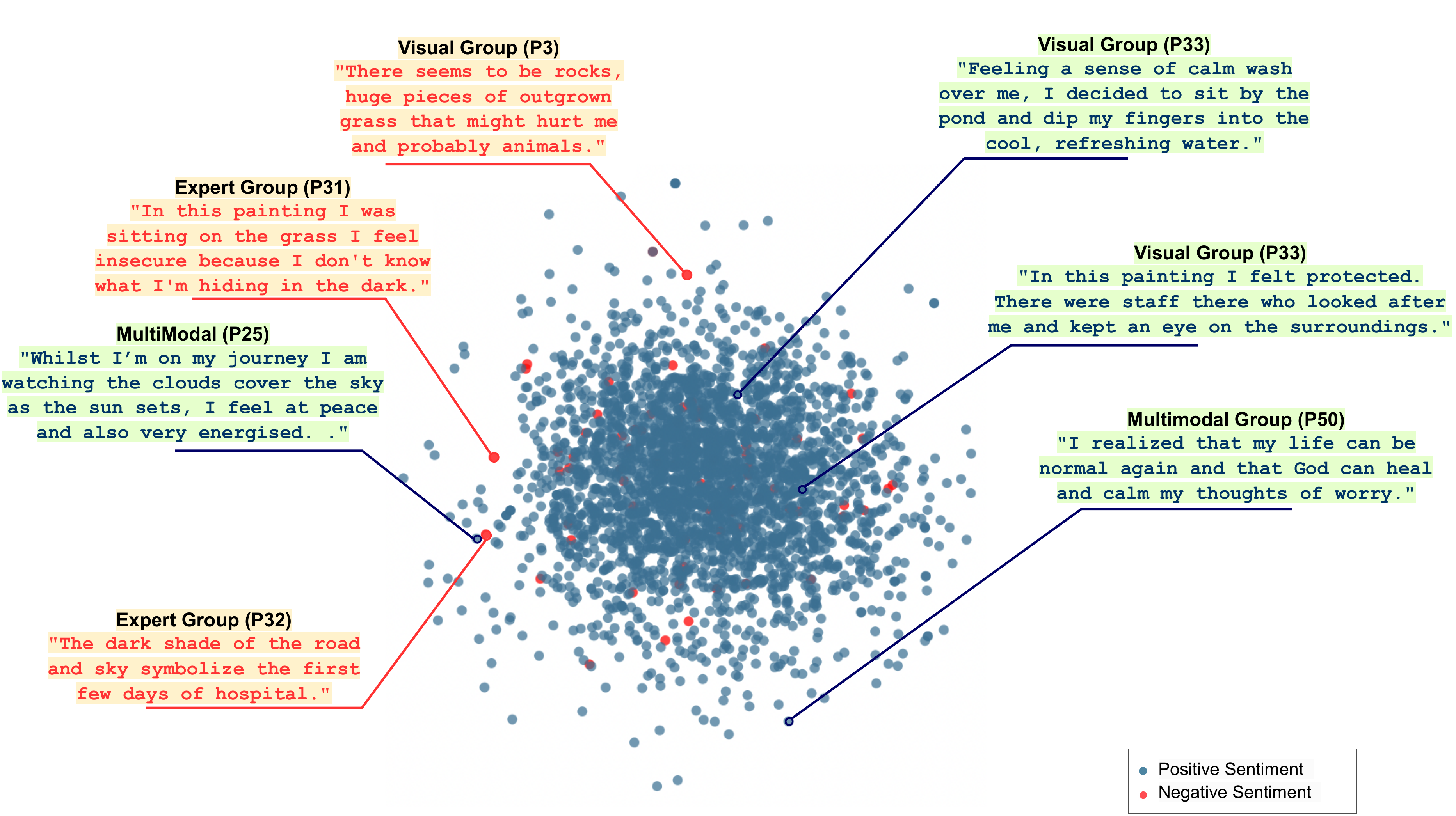}

    \caption{
       t-SNE projection of sentence embeddings from user reflections
    }
    \label{fig:embeddings_sentiment}
\end{figure*}

We conducted sentiment analysis on the reflections of participants to gain deeper insights into their experience. 
We used the pre-trained transformer-based sentiment analysis model \texttt{bert-large-uncased-sst2} 
from the Hugging Face Transformers library.\footnote{https://huggingface.co/models} 
This model is a  fine-tuned version of \texttt{bert-large-uncased} which was trained on the Stanford Sentiment Treebank v2 (SST2)\footnote{https://nlp.stanford.edu/sentiment}; part of the General Language Understanding Evaluation (GLUE) benchmark\footnote{https://gluebenchmark.com}. It is well-suited for a wide range of NLP tasks due to its large size and general language understanding capabilities. It has also demonstrated exceptional success in sentiment analysis tasks.  Leveraging this model the result of our sentiment analysis, illustrated in \autoref{fig:sentiment} indicates overwhelmingly positive sentiments expressed by participants 
in response to their interaction with the recommended paintings. 
However, it is noteworthy that a subtle trend emerged within the expert group, 
where a marginal 4\% of sentences conveyed negative reflections, 
while sentences from the Visual group exhibited a slightly lower 2\% negativity rate. 
In stark contrast, the sentences from the Multimodal group displayed an absence of negative sentiments altogether, 
echoing the similar pattern observed in the mood change PANAS, 
as well as recommendation quality measures.  
\autoref{fig:embeddings_sentiment} shows some sample sentences from the positive and negative reflections of participants per group on a 2D projection map of all reflection sentences using the non-linear projection t-SNE algorithm~\cite{Maaten:2008:tSNE}.
 
\begin{table*}[!h]
\small
\begin{tabularx}{\textwidth}{p{0.15\linewidth}X<{\arraybackslash}p{0.55\linewidth}X}
\toprule
\textbf{Themes}  & \textbf{Theme Description} & \textbf{Example Quotes} \\
\midrule
Hope and Purpose  & Drawing one's attention toward more positive prospects, reminding them of hope and purposefulness. & \texttt{“The style of the painting makes it more dynamic and the colors also make it more appealing. ... This color palette really brightens my mood and puts me in a better place mentally. ... I felt happy and hopeful about the future. I was in a very good mood.”} \textbf{P07}
\\
\midrule
Rejuvenation & Supporting one to feel recharged through a sense of being carefree, calm, and relaxed. & \texttt{“In this painting, I felt calm, relaxed, and refreshed. ... I felt my spirit being re-energized as I cast away my concerns and worries.”} \textbf{P21} \\
\midrule
Engagement  & Supporting one to immerse into the visual art by triggering one’s attention and interest. & \texttt{“... the shadows and the light are done perfectly to give me that feeling of being there, actually seeing myself walking down the road on my way to the dam. This is a place I know in my mind, thus making the painting very engaging to me.”} \textbf{P35} \\
\midrule
Safety  & Promoting a sense of safety through elements that signal a safe and familiar environment. & \texttt{“It (the scenery in the painting) is very comforting and close to home. The subject matter is also important because it represents a situation where I'm in someone's company. ... I felt safe and at ease because I'm in good company and surrounded by nature.”} \textbf{P04}
\\
\midrule
Sensory Pleasure & Providing pleasure through rich sensory stimulation that either directly comes from visual art or is derived from memory through visual triggers. & \texttt{“In this painting, I found myself standing on the edge of a serene, sun-kissed meadow, surrounded by vibrant wildflowers swaying gently in the breeze. The colors were so vivid that I could practically feel the warmth of the sun on my skin and the softness of the grass beneath my feet. ... I began to meditate, allowing the beauty of the painting to fill my senses.”} \textbf{P42} 
\\
\midrule
Relevance & Presenting subject matter relevant to one’s specific situation that can stimulate memories or constructive reflections. & \texttt{“The scenery and style helped me remember previous times I have been in this situation. It made me think of all the good times and look forward to more good times. It was a very positive way to pass the time and think about what I need to be focusing on in life.”} \textbf{P18}\\
\midrule
Personal Preference & Increasing pleasure with visual stimulation that meets one’s preference. & \texttt{“... since I love to visit places with a lot of forest, where there is not much noise and if there is noise, it is of nature, this painting helped my experience a lot.”} \textbf{P74} 
\\
\bottomrule \\
\end{tabularx}
\caption{Themes and Example Quotes from Some Participants}
\label{tab:themes}
\end{table*}
While the sentiment analysis offered valuable insights, it's important to acknowledge that our pre-trained sentiment analysis model based on the general-purpose language model BERT, may not capture all sentiment subtleties, particularly those related to healing.  Thus, to account for this we further conducted a qualitative evaluation through reflexive thematic analysis (RTA)  \cite{braun2006using} to identify healing elements in the recommended arts. The participant quotes were reviewed and labeled through open coding. These codes captured information related to the elements that contribute to patients’ affective states. The results of open coding were re-analyzed to identify important and general concepts. These concepts were interpreted and categorized into a total of seven themes based on how they contribute to eliciting positive emotions and moods: hope, rejuvenation, engagement, safety, sensory pleasure, relevance, and personal preference (see \autoref{tab:themes} for the list of themes and example quotes). The identified themes show different elements that contribute to the healing of individuals, each leading to unique paths to healing. We observed that these paths involve symbolic associations, such as hope and purpose for some individuals, while for others, they involve aesthetic values, such as sensory pleasure or personal preference. Therefore, the list of these themes can be used as elements to select healing paintings tailored to the needs of individuals: one who is drained might seek rejuvenation, while one who feels insecure might seek safety or familiarity. These findings underpin the importance of personalization of healing art for therapeutic purposes. Most identified themes, such as engagement, safety, sensory pleasure, familiarity, and personal preference, echo the relaxing visual nature elements found in other studies \cite{kim2023relaxation, appleton1975experience, wilson1986biophilia}. Hope and purpose are often emphasized as core elements in one’s coping and healing process \cite{allen2008coping}. This analysis also revealed the significance of including these elements in art to function as a healing mediator. We observed that the absence of such elements had led to rather negative affective experiences in all three groups (Expert, Visual, and Multimodal groups). An absence of engagement, for instance, contributed to boredom: \emph{“This one just doesn’t draw me in, (…) I would find it very boring and not part of any healing experience“} (participant 54 from Visual Group). Likewise, an absence of hope led to feeling tense: \emph{“This image doesn't really inspire hope.  Instead, it brings pessimistic thoughts. I hope for a more uplifting and positive atmosphere during my time here”} (Participant 122 from Multimodal group). An absence of safety led to feeling gloomy: \emph{“This painting evokes the feeling of being lost. I felt like there was nowhere to hide so I just had to face whatever feeling I had”} (Participant 25 from Expert group).

\section{Discussion}


Based on our findings, we can answer positively our research question posed at the beginning of this paper.
That is, VA RecSys algorithms can indeed support the rehabilitation of post-ICU patients using art therapy.
This has important implications in several fronts, as we discuss below.

\subsection{Personalised visual art as PICS intervention}
\label{subsec:discuss_one}

We have explored the potential of art therapy as a PICS intervention and have tested VA RecSys as a means to personalize visual art for this goal. Overall, we found that this comparatively new approach to using art, which combines narrative techniques with personalized recommendations, allowed participants to engage with various healing elements in the artworks. Additionally, our findings show that personalized guided art therapy is effective in temporarily alleviating negative emotions and enhancing positive emotions as well as enhancing mood states. This suggests that by increasing its duration and dosage, it has the potential to address the psychological aspects of PICS as an intervention, which could potentially result in more lasting effects in enhancing the affective state of patients. Importantly, we utilized nature-based artwork in this study. While the use of nature-based artwork to support former ICU patients is a novel approach, previous studies have demonstrated the therapeutic effects of nature-based visuals in various forms, ranging from a real nature view \cite{ulrich1984view} to static as well as dynamic versions of virtual nature \cite{hung2022exploring, malbos2013creation,veling2021virtual}. The results of our study contribute to the ongoing research efforts in applying nature-based visuals for therapeutic purposes, demonstrating their potential to support the healing process of patients and enhance their psychological well-being. Furthermore, in line with a recent study \cite{kim2023relaxation} that has highlighted the influence of personal characteristics on the impact of visual nature experiences, our study suggests the potential for a higher level of personalization with the support of RecSys.

Finally, we should mention that the process of guided art therapy in this study engaged an expert solely during the preparation phase, remaining independent throughout. This suggests the potential for developing guided art therapy as an intervention for remote and self-administered use, which could help alleviate the primary constraints of current PICS interventions, known for their high demand on healthcare professionals.

\subsection{Crossing boundaries: VA Recsys - From entertainment to therapy}
\label{subsec:discuss_two}
VA RecSys engines originally emerged as a means to enhance user experience in the entertainment field. Particularly, recent approaches boosted by machine learning techniques have undoubtedly demonstrated their potential in supporting users such as museum visitors and art enthusiasts to discover art pieces that are tailored to their personal preferences and interests. Furthermore, their ability to uncover complex semantics embodied within visual art made them powerful tools to support learning and discovery by exposing users to novel content.  While the art entertainment industry has benefited from these advancements, our study sheds light on the remarkable potential of VA RecSys to transcend the space of entertainment, emerging as therapeutic tools within the healthcare domain. In the field of art entertainment, users seek diversion, enjoyment or relaxation while service providers strive to not only enhance user engagement and satisfaction but also drive up revenue. VA RecSys has long been at the forefront, seamlessly aligning these dual objectives.

In stark contrast to entertainment, therapy serves a deeper purpose; it is a journey of healing and self-discovery. The use of art in therapy has been proven to provide individuals with a unique medium to express complex emotions, confront traumatic experiences, and embark on the path to recovery. Art therapy, in particular, has emerged as a powerful tool in the hands of trained professionals to address a wide range of psychological and emotional challenges. The key to effective therapy lies in personalization and relevance. Patients seek a therapeutic experience that resonates with their individual needs and experiences. Thus, a careful selection of paintings tailored to the individual patient speaks to their unique circumstances, fostering self-reflection and healing. By introducing VA RecSys to the domain of therapy, we have showcased its remarkable potential to assist professionals not just in curating personalized artworks from a vast selection but also in delivering precise, tailored treatment to patients.

The intent here is not mere engagement as in the domain of entertainment but rather the transformation of the individual's affective state and well-being. Thus, the adoption of VA RecSys algorithms from entertainment to the context of therapy requires rigorous quality control before being deployed in a system facing patients. As informed by our pilot test in \autoref{subsec: pilot} not all top-performing VA RecSys engines in the entertainment domain were found to be appropriate for the purpose of therapy. Particularly recommendations from our text-based engines BERT and LDA tend to contain paintings that feature contents evoking negative emotions and with potentially harmful consequences. Therefore, we need to underscore the importance of acknowledging the risks involved and taking the necessary precautions when adopting these algorithms. On the contrary, our image-based and fusion-based engines produced paintings that were deemed appropriate by experts and our results also indicate that they were even perceived to support healing better than expert-curated recommendations. 
However, the promising results we observed may indicate a potential for a Human-in-the-Loop approach wherein experts fine-tune the recommendations generated by VA RecSys engines. While experts play a crucial role in ensuring the quality of paintings, VA RecSys could significantly reduce their workload (e.g., sifting through thousands of individual paintings from a database), which is reportedly a concern \cite{pastores2019workforce, chang2019relationship}, thereby enhancing the potential for scaling up guided art therapy to bring benefits to more patients. This is nonetheless an exciting opportunity of Human-AI collaboration for future work.

\subsection{Looking ahead: Potential of VA RecSys in healthcare beyond PICS intervention}
\label{subsec:discuss_three}
\begin{figure*}[!ht]
    \centering
    \def\h{3.4cm}
    \subfloat[]{\includegraphics[height=\h]{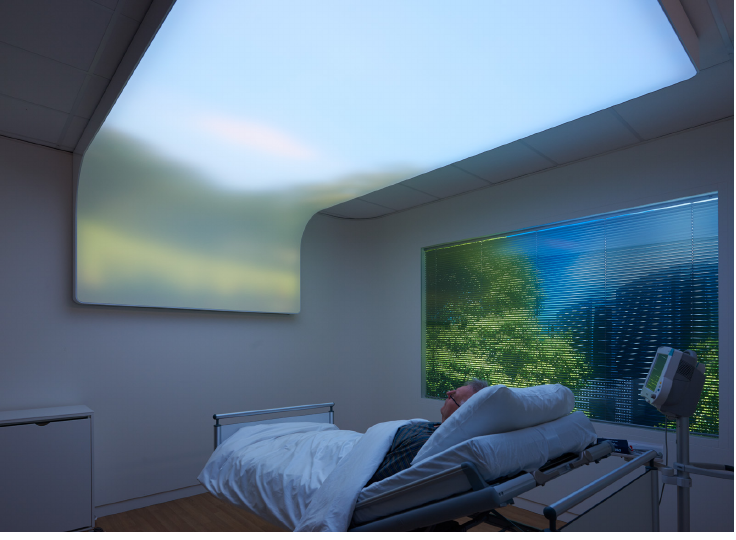}} \hspace{0.3cm} 
    \subfloat[]
    {\includegraphics[height=\h]
    {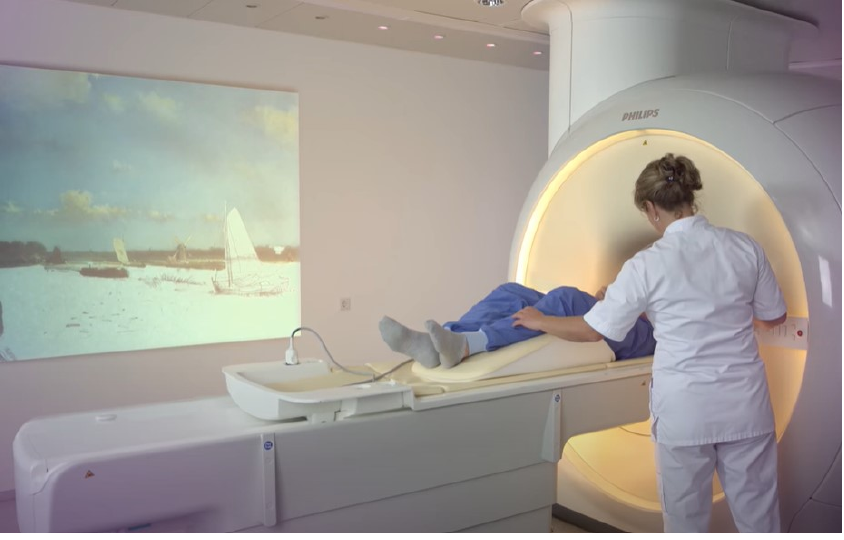}} \hspace{0.3cm} 
     \subfloat[]
    {\includegraphics[height=\h]
    {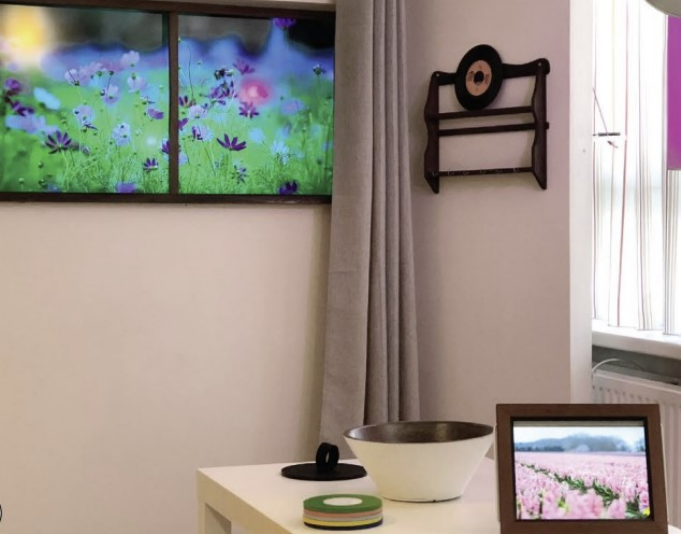}}
    \caption{VA RecSys enabled Art-therapy in ICU and beyond. (a)	ICU with a luminous LED ceiling by Philips VitalMind.\protect\footnotemark{} (b) MRI room with a projection by Philips Ambient Experience\protect\footnotemark[12]{} (c) Virtual window and digital frame for Residential Dementia Care, AmbientEcho by Thoolen et al. 2020 \cite{thoolen2020ambientecho}
    }
    \label{fig:ICU_art}
\end{figure*}

Our exploration of VA RecSys in the context of PICS is but a glimpse into the vast potential of this innovation within healthcare. Particularly, in light of our promising results observed in PICS treatment, one natural extension of this approach is to implement VA RecSys-assisted visual art into the ICUs (see \autoref{fig:ICU_art}-a). This could support PICS prevention and the well-being of patients by providing essential emotional support (e.g., reducing fear and anxiety) through recommending personalized art.  

The adaptability of VA RecSys-assisted art therapy holds promise in areas far beyond the boundaries of intensive care where the use of visual content as a positive distraction is already active. For instance, the use of projection creates a more relaxing experience in an MRI room where patients can get easily worried and feel discomfort (see \autoref{fig:ICU_art}-b). In the context of residential care, as another example, a virtual window or digital frames are used to support cognitive activation and recovery (see \autoref{fig:ICU_art}-c). The implications of our findings extend the potential of VA RecSys engines in the intersection of AI and healthcare. The role of VA RecSys engines in facilitating the use of visual art as a positive distraction is merely the tip of the iceberg, hinting at a future where technology enables more holistic and personalized care, amplifying our capacity to heal and connect on a profound level. 

\footnotetext{Source: \url{www.philips.nl}}

\section{Limitations and future work}
While our study highlights the potential of VA RecSys within a therapeutic context, 
we acknowledge certain limitations and chart out promising directions for future research. 
Firstly, we have observed significant disparities when using VA RecSys in therapeutic context compared to its conventional application in entertainment. While our study has shed light on these distinctions, it has also highlighted risks associated with therapeutic use, especially when leveraging text-based models. This may partially be attributed to the quality of the text data source. 

Data quality, underpinning VA RecSys recommendations, plays a pivotal role in its effectiveness. We have employed artist-curated descriptions of 2,368 paintings  from the National Gallery dataset, but it is evident that these descriptions may not fully encapsulate the intricate affective attributes of the artworks. Thus, improving these models for therapeutic purposes can potentially be achieved by curating richer, more comprehensive affective descriptions. Although this entails substantial content curation efforts and a thorough evaluation, we believe it is a worthwhile endeavour. For future work it would also be beneficial to implement tree-based indexing data structures to scale up more efficiently for larger datasets.

Another limitation is that our current preference elicitation method relies on users selecting a single preferred painting, which may oversimplify their preferences. An area ripe for improvement involves allowing users to rate all the sample paintings that provide comprehensive representations of affective states (i.e., calmness, restoration, and cheerfulness), thereby capturing to what extent they resonate towards each affective dimension. By using these ratings as weights and projecting them into the embedding space, we can refine recommendation accuracy and granularity. Thus, the development and evaluation of VA RecSys combining the curation of high-quality data with such preference weighting mechanisms holds potential to improve the current approach in therapeutic settings. Additionally, by deriving more personalised content recommendations that uncover deeper semantics of artworks, this may also extend current VA RecSys approaches mostly limited in entertainment~\cite{Yilmaleiva23, yilma2020personalised} to benefit other areas such as education, blend learning, and discovery of artistic concepts.

As hinted in the above subsection, one particularly promising avenue is the exploration between humans and AI systems in the context of therapy.  Here, experts can fine-tune VA RecSys-generated recommendations to align them precisely with individual patient needs. Investigating the dynamics of such collaborative efforts and developing tools to facilitate expert interventions could significantly enhance the therapeutic value of VA RecSys. This exciting direction opens doors to more targeted and personalized therapy experiences, bridging the gap between technology and human expertise.
 Furthermore, while we have gained valuable insights with the current sample (i.e., former patients with psychological symptoms of PICS), validation with patients exhibiting PICS symptoms is warranted.
Finally, one key challenge lies in comprehending the reasoning behind the VA RecSys recommendations, which remains a critical aspect in determining model performance in different contexts.  The explanation of machine learning models has been a longstanding challenge in the field of AI. Nevertheless, recent strides have been made in the realm of explainable AI, with emerging techniques and methodologies. Leveraging these innovative approaches to provide more transparent and interpretable explanations for VA RecSys recommendations holds substantial promise. This advancement can facilitate a Human-in-the-Loop approach, empowering experts to refine and enhance therapy efforts with greater precision.

\section{Conclusion}

We have studied Machine Learning-based VA RecSys approaches to enable personalized therapeutic visual art experiences for post-ICU patients. We have evaluated the relevance of the recommendations for therapeutic purposes as compared to expert-curated recommendations. Our results suggest that Visual and Multimodal VA RecSys engines compare favourably with expert-curated recommendations, indicating a great potential to support the delivery of personalised and targeted art therapy for PICS prevention and treatment. Overall, our study marks a significant step towards integrating VA RecSys in the context of therapy. Considering future research directions, this work points to the exciting potential for further advancements in the field of AI-assisted therapy and recommendation systems.
The implications of our findings extend to patient-centred care, early intervention, and health promotion.

\begin{acks}
We thank Thomas Falck, MSc (Philips) and Dr. Esther van der Heide (Philips) for their advice on PICS research, and Prof. Dr. Geke Ludden (University of Twente) and Dr. Thomas van Rompay (University of Twente) for their advice on developing guided art therapy used in this study. We extend our thanks to the participants of our study for sharing their valuable experiences, and to the anonymous reviewers for their constructive comments.
This work was supported by the Horizon 2020 FET program of the European Union through the ERA-NET Cofund funding grant CHIST-ERA-20-BCI-001
and the European Innovation Council Pathfinder program (SYMBIOTIK project), and the Top Technology Twente Connecting Industry program (TKI Topsector HTSM), which is partially funded by Philips. 
\end{acks}

\bibliographystyle{ACM-Reference-Format}
\bibliography{refs}

\end{document}